\documentclass[numbook,natbib,runningheads]{svjour3}
\usepackage{graphicx,mathptmx,color}

\newcommand{\mr}{\ensuremath{\mathrm}}

\newcommand{\aap}{    {\rm Astron. Astrophys.}\ }

\newcommand{\aapr}{   {\rm Astron. Astrophys. Rev.}\ }

\newcommand{\apj}{    {\rm Astrophys. J.}\ }
\newcommand{\apjl}{   {\rm Astrophys. J. Lett.}\ }
\newcommand{\apjs}{   {\rm Astrophys. Suppl.}\ }

\newcommand{\araa}{   {\rm ARAA}\ }

\newcommand{\grl}{    {\rm Geophys. Res. Lett.}\ }

\newcommand{\jgr}{    {\rm J. Geophys. Res.}\ }
\newcommand{\mnras}{  {\rm Mon. Not. Roy. Astron. Soc.}\ }

\newcommand{\pasp}{   {\rm Pub. Astron. Soc. Pac.}\ }
\newcommand{\pasj}{   {\rm Pub. Astron. Soc. Japan}\ }

\newcommand{\solphys}{{\rm Solar Phys.}\ }

\usepackage{makeidx} \makeindex
\makeindex

\begin{document}

\title{Microflares and the Statistics of X-ray Flares}

\author{I.~G.~Hannah$^{1}$, H.~S.~Hudson$^{2}$, M.~Battaglia$^{1}$,
S.~Christe$^{3}$, J.~Ka\v{s}parov\'{a}$^{4}$, S.~Krucker$^{2}$, M.~R.~Kundu$^{5}$, and A.~Veronig$^{6}$}

\institute{$^{1}$School of Physics \& Astronomy, University of
Glasgow, Glasgow, G12 8QQ, UK\\
$^{2}$Space Sciences Laboratory, University of California at Berkeley, Berkeley,
CA, 94720-7450, USA\\
$^{3}$NASA Goddard Space Flight Center, Greenbelt, MD 20771,
USA\\
$^{4}$Astronomick\'{y} \'{u}stav AV \v{C}R, v.v.i., Fri\v{c}ova 298,
Ond\v{r}\v{e}jov, 251 65, Czech Republic\\
$^{5}$ Dept. of Astronomy, Univ. of Maryland, College Park, MD 20740\\
$^{6}$Institute of Physics/IGAM, University of Graz, Universit\"{a}tsplatz 5,
8010, Graz, Austria}

\date{}

\titlerunning{Microflares and Flare Statistics}\authorrunning{Hannah et al.}
\maketitle
\begin{abstract}
This review surveys the statistics of solar X-ray flares, emphasising the new views
that \textit{RHESSI} has given us of the weaker events (the microflares).
The new data
reveal that these microflares strongly resemble more energetic events in most
respects; they occur solely within active regions and exhibit high-temperature/nonthermal 
emissions in approximately the same proportion as
major events. 
We discuss the distributions of flare parameters (e.g., peak flux) and
how these parameters correlate, for instance via the Neupert effect. We also
highlight the systematic biases involved in intercomparing data representing
many decades of event magnitude. The intermittency of the flare/microflare
occurrence, both in space and in time, argues that these discrete events do not
explain general coronal heating, either in active regions or in the quiet Sun.
\end{abstract}


\tableofcontents

\section{Introduction}

A solar flare is a rapid and transient release of energy in the solar corona,
associated with electromagnetic radiation from radio waves to $\gamma$-rays,
local plasma heating to tens of MK, particles accelerated to~GeV, violent mass
plasma motions, and shock waves. In the largest events, a spectacular variety of
phenomena can be studied in detail across many wavelength ranges to try to
understand the processes involved \citep[see examples in ][]{chapter2}. 
The X-ray
emission observed is of particular interest as it shows the accelerated particles
and intense heating, and therefore provides the most direct insights into the
physics of the basic energy release. Hard X-rays (HXRs), from about 10~keV to
hundreds of~keV, are primarily produced via thick-target 
bremsstrahlung\index{bremsstrahlung!thick-target}
\citep{1971SoPh...18..489B,chapter7} in which the coronal accelerated electrons
are stopped instantaneously through Coulomb collisions\index{Coulomb collisions} with denser material in the lower solar atmosphere. 
Soft X-rays (SXRs), typically the component below tens of keV, are thermal emission (lines and continua) by plasma of a few to tens of MK. 
This emission results to some extent from heating at the site of energy
release, but mainly from new coronal material evaporated from the
chromosphere during the impulsive phase \citep[see][]{chapter2}. Given the
wealth of X-ray flare data, it is possible to study large samples of these events,
allowing the statistics of the events to provide clues to the underlying processes
behind the emission. This article reviews such flare studies, the results,
interpretation and limitations, from a predominantly X-ray viewpoint. In
particular, we discuss the advances that have been made in extending this analysis
to weaker HXR events (\emph{microflares}) using the \textit{Reuven Ramaty High Energy
Solar Spectroscopic Imager (RHESSI)} \citep{2002SoPh..210....3L}.
\index{satellites!RHESSI@\textit{RHESSI}}\index{RHESSI@\textit{RHESSI}}

\begin{figure}
\centering\sidecaption
\includegraphics[width=0.7\columnwidth]{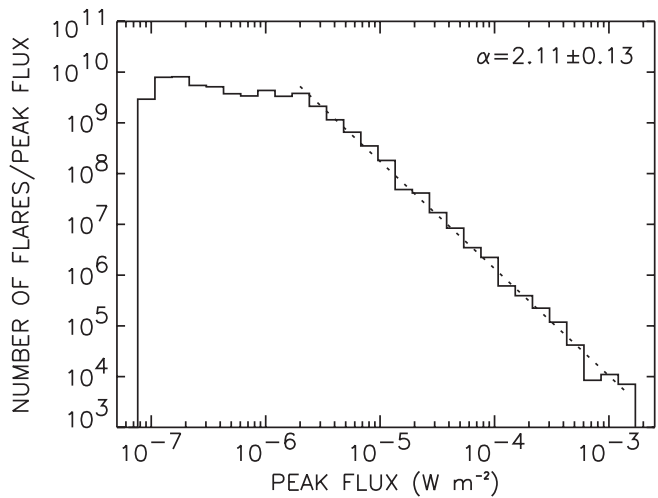}
\caption{\label{fig:hannah_goesdist}Frequency distribution of the peak
1-8~\AA~ SXR \textit{GOES} flux of 49,409 flares between 1976 and 2000
\citep{veronig2002}. Note that this study did not include the A-class
events ($<$10$^{-7}$~W~m$^{-2}$). In addition, there was no attempt at
background subtraction, so the resulting value of~$\alpha$ should
be regarded as an upper limit (background predominantly affects
smaller flares, artificially steepening the distribution).}
\end{figure}

One of the most remarkable features of flare surveys is the range of magnitudes
these transient phenomena span. The observed~SXR flux extends over five orders
of magnitude (see Figure~\ref{fig:hannah_goesdist}) and over nine orders of
magnitude in energy (see Figure~\ref{fig:hannah_eng}; we return to discuss these
Figures in detail in Section~\ref{sec:hannah_obsdist}). 
The observations come from many spacecraft: e.g., \textit{Orbiting Solar Observatory (OSO-3)} and \textit{(OSO-7)}, the \textit{Solar Maximum Mission (SMM)}, the \textit{International Cometary Explorer (ICE, a.k.a. ISEE-3)},  the \textit{Compton Gamma-Ray Observatory (CGRO)}, \textit{Yohkoh}, the \textit{Solar Heliophysical Observatory (SOHO)}, \textit{GRANAT}/ WATCH, and the \textit{Transition Region and Coronal Dynamics Explorer (TRACE)}.
\index{satellites!CGRO@\textit{CGRO}}
\index{satellites!SMM@\textit{SMM}}
\index{satellites!ICE@\textit{ICE}}
\index{satellites!ISEE-3@\textit{ISEE-3}}
\index{satellites!TRACE@\textit{TRACE}}
\index{satellites!Yohkoh@\textit{Yohkoh}}
\index{satellites!SOHO@\textit{SOHO}}
\index{satellites!OSO-3@\textit{OSO-3}}
\index{satellites!OSO-7@\textit{OSO-7}}
\index{satellites!ICE@\textit{ICE}}
\index{satellites!GRANAT@\textit{GRANAT}!WATCH}
The distributions of these
and other flare parameters, such as emission at other wavelengths and for other durations,
have the further striking property that they all can be well represented by a power law of the form
\begin{equation}
\label{eqn:hannah_powlawdef} f(x;\alpha)=Cx^{-\alpha},
\end{equation}
where $f$ is the probability density function (PDF; this is often called the flare
frequency distribution)\index{flare frequency distributions} of the flare parameter $x$, $\alpha>0$ is the power-law
index and $C$ is a scaling constant\index{probability distribution function}.\index{frequency!occurrence}
In the case of the flare's energy $U$, the
quantity $f(U)dU$ is the fraction of events per unit time releasing energy between
$U$ and $U+dU$. This power-law nature of the frequency distribution has long
been observed, first noticed in solar radio bursts by \citet{1956PASJ....8..173A}.
The fact that so many of the flare characteristics have such a distribution is thought
to arise from the corona being in a self-organized critical state\index{self-organized critical state}\index{flare models!self-organized criticality (SOC)}
\citep{1991ApJ...380L..89L}, discussed further in Section~\ref{sec:hannah_whyapowlaw}.
Here, the underlying energy release from the largest to the smallest flares occurs as an
``avalanche,'' similar to the theory for earthquake occurrence or stress relief in
general \citep{1986JGR....9110412K}.
\index{nanoflares!Parker's hypothesis}
These properties have brought forth interesting interpretations. The power-law
distribution means that small flares are more numerous. 

Could these weaker events at the currently unobservable ``nanoflare''\index{nanoflares} 
magnitude level (about
$10^{-9}$ times the energy of a large flares) directly constitute the hot corona,
thus explaining its heating\index{coronal heating!nanoflares}? 
This is often called Parker's nanoflare mechanism\index{Parker's nanoflare mechanism}
\citep{1988ApJ...330..474P}. The answer to this question requires knowledge of
the total power that is contained in the flare distribution\index{flare frequency distributions!total power}, i.e.

\begin{equation}\label{eqn:hannah_totpow}
P=\int_{U_\mr{min}}^{U_\mr{max}}
f(U;\alpha)UdU=\frac{C}{2-\alpha}\left[{U_\mr{max}}^{2-\alpha}-
{U_\mr{min}}^{2-\alpha } \right ].
\end{equation}

\noindent Since $U_\mr{min} \ll U_\mr{max}$ by definition, we can see that in the
case $\alpha>2$ in Equation (\ref{eqn:hannah_totpow}) the low-energy half of the
distribution would contain the most energy \citep{1991SoPh..133..357H}. In this
situation the smallest flares fit the requirement for heating the corona as they
have a high occurrence rate and release more net energy than the large flares.
Therefore, to understand the role of flares in coronal heating, the power-law index
$\alpha$ has to be accurately estimated; this is problematic as the energy is
inferred from observations and thus subject to potentially large errors and biases.
In addition, Equation~(\ref{eqn:hannah_totpow}) assumes that the distributions
continue into the unobservable low-energy range. Even if the deduced $\alpha$
were large enough (i.e., larger than two),  that would not automatically mean that
small flares heat the corona. For further proof we would require observations
showing the power-law index to maintain its value down to the distribution's
physical limit\index{flare frequency distributions!physical range}, rather than its observational limit\index{flare frequency distributions!observational range}.

\index{GOES@\textit{GOES}!illustration}
\index{satellites!GOES@\textit{GOES}}

Between the largest flares and these theoretical nanoflares lie the microflares
(nominally with energies of order $10^{-6}$ times those of large flares), and it is
with these events that \textit{RHESSI}'s HXR observations have made major advances.
\textit{RHESSI}'s continuous spectral coverage, down to 3~keV, uniquely allows the study of
the transition between the thermal and nonthermal emissions with the same
instrument.
\index{hard X-rays!RHESSI@\textit{RHESSI}!microflare observations}
In particular, \textit{RHESSI}'s view of their nonthermal characteristics has
allowed us to study smaller-scale active region events with simpler structures
than the major events that often attract the most attention. Perhaps these can
help us to isolate the essential physics, which could then be applied in more
complex situations.\index{active regions!simple structures}

We begin this article with a general overview of flares and flare-like
brightenings across the range of magnitudes (Section~\ref{sec:hannah_fc}). This
discloses similar properties between major and minor events, but systematic
differences do occur. In Section~\ref{sec:hannah_obsdist}, we learn what we can
from these differences about the physics, taking advantage of the very large
numbers of events in major surveys, including \textit{RHESSI}'s new HXR views. We also
discuss in this section the biases that arise in the statistical surveys and possible
methods for obtaining the unbiased intrinsic distributions. In
Section~\ref{sec:hannah_whyapowlaw} we briefly discuss how the power-law
nature of the flare parameters arises. Conclusions and discussions are given in
Section \ref{sec:hannah_cons}.

\section{From major to minor flares}\label{sec:hannah_fc}

\subsection{Flare classification \& general properties}
\index{flares!classification}\index{flare classification}

The most powerful ordinary flares have energies estimated at above
10$^{33}$~ergs and present a spectacular range of phenomena, easily observed
across the wavelengths. The first flare observed was a powerful  event in~1859,
detectable through its  small, intense white-light emission patches\index{white-light flares!emission patches} as described
by \cite{carrington1859} and corroborated by \cite{hodgson1859}. Remarkable
terrestrial effects accompanied this flare and also followed it after an interval of
half a day. 
This event anticipated much of the complexity of flares as we know
them today, but it was not until the 1940s that ``flare'' was accepted as the term to
describe these transient phenomena \citep{newton1943,richardson1944}. Events
with total energy about a millionth smaller than large flares (about $10^{27}$~erg),
became known as ``microflares'' \citep{1983SoPh...89..287S,1984ApJ...283..421L}.
\index{microflares!hard X-ray!discovery}
\index{Parker's nanoflare mechanism}\index{nanoflares!energy range}
Parker hypothesized that even smaller flares, ``nanoflares,'' with energies of order
one billionth of large flares or about $10^{24}$~erg, could be the basic unit of a
localized impulsive energy release \citep{1988ApJ...330..474P}.

\begin{figure}
\centering\sidecaption
\includegraphics[width=0.7\columnwidth]{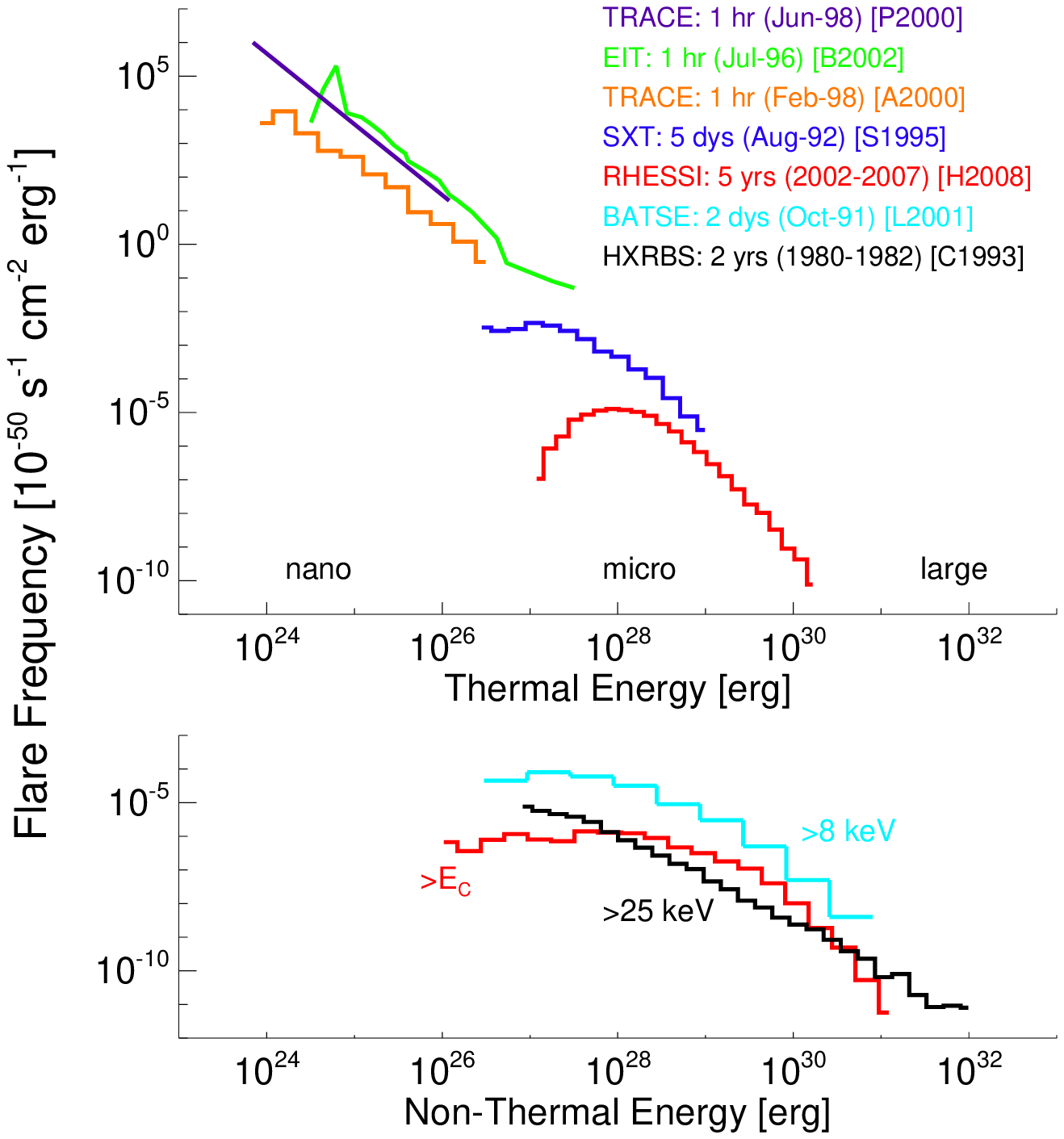}
\caption{
\label{fig:hannah_eng}The energy distributions for solar flares. The nonthermal
energy distribution is shown for large flares $>$25~keV observed with
\textit{SMM}/HXRBS \citep{crosby1993}, microflares $>$8~keV from
\textit{CGRO}/BATSE  \citep{lin2001} and microflares $>E_\mr{C}$ (above the
low energy cutoff) from \textit{RHESSI} \citep{hannah2008}. The thermal energy
distribution is shown for microflares with \textit{RHESSI} \citep{hannah2008}
and \textit{Yohkoh}/SXT \citep{shimizu1995} and EUV nanoflares with \textit{TRACE}
\citep{parnell2000,asch2000} and \textit{SOHO}/EIT 
\citep{benz_krucker2002}. This figure is deceptive as it is comparing energy
distributions of different flare energy components, each involving different
instrument and selection effects, and were obtained over different periods of
different solar cycles.}
\end{figure}
\index{flare frequency distributions!illustration}

Quantitative flare classification is based on the 1-8~\AA~SXR flux observed by \textit{GOES}\index{satellites!GOES@\textit{GOES}}\index{satellites!SMM@\textit{SMM}!HXRBS}\index{satellites!CGRO@\textit{CGRO}!BATSE}\index{satellites!Yohkoh@\textit{Yohkoh}!SXT}.
Large flares have X$n$-class, indicating a peak flux of
$n\times10^{-4}~\mr{W~m}^{-2}$, the largest events being above X10.
\index{GOES@\textit{GOES}!classification}
This classification decreases through the decades of M, C and B-class flares down to the
smallest A$n$-class events with $n\times10^{-8}$~W~m$^{-2}$ and the sensitivity
limit of the detector. The classification of flares and the associated range of \textit{GOES}
fluxes is shown in Table~\ref{tab:hannah_class}. 
The largest \textit{GOES} flare was SOL2003-11-04T19:53 (X17.4), which saturated the detectors at
$18\times10^{-4}~\mr{W~m}^{-2}$ (i.e., class X18). 
\index{flare (individual)!SOL2003-11-04T19:53 (X17.4)!most energetic event}
It is estimated that this flare was an X28 event with range X25 to X31 \citep{2004AAS...204.4713K}. 
In terms of \textit{GOES} classification, the Carrington flare\index{flares!Carrington} 
appears to have been a large soft X-ray flare of magnitudfe
$>$X$10$ \citep{2003JGRA..108.1268T} and like many large flares resulted in a
major geomagnetic storm \citep{2004SoPh..224..407C}. 
\index{flare (individual)!SOL1859-09-01T11:18 (pre-\textit{GOES})}
This region also produced
one of the largest solar energetic proton fluences at the Earth in the last 500 years
\citep{2001ICRC....8.3209M}. 
These most energetic flares occur in active regions,
often when new flux emerges into an already-complex magnetic structure.\index{active regions!association with major flares}
From the start of the \textit{GOES} observations in 1975, to the start of cycle 24 in 2009,
22 flares $>$X$10$ had been observed. 
In total, 359 \textit{GOES} X-class flares, 4708
M-class, 32784 C-class and 11558 B-class flares occurred between 1976 and 2000
\citep{veronig2002}. Although more small flares are expected due to the inherent
power-law nature of the flare distribution, they are hidden by the higher
background during active times. The majority of large flares seem to occur during
the peak of the solar cycle heading into the decay phase, though this behavior
can vary dramatically from cycle to cycle \citep[e.g.,][]{2007ApJ...663L..45H}. These
two features can be seen in the \textit{GOES} flaring rate shown in Figure~\ref{fig:hannah_mftime}, discussed in Section \ref{sec:hannah_microar}.

\begin{table}\begin{centering}
\caption{\label{tab:hannah_class}Different flare classifications and the associated
ranges of \textit{GOES} flux (SXRs) and HXRs.}
\begin{tabular}{p{2.cm}p{3cm}p{1.cm}p{0.75cm}p{1.5cm}p{2cm}}
  \hline\noalign{\smallskip}
  Flare Size & Description & Energy [erg]&\textit{GOES} Class & \textit{GOES} Flux [W~m$^{-2}$]&
HXR emission \\
  \hline\noalign{\smallskip}
  Large$\setminus$normal &Active region phenomena &$\leq 10^{33}$ &C,M,X &
  $10^{-6}-10^{-3}$ & $>$25~keV \\
    Micro$\setminus$ARTB$^1$ &Active region phenomena &$\sim 10^{27}$ &A,B &
    $10^{-8}-10^{-6}$ & $10-30$~keV \\
      Nano&Unobserved basic unit of localized impulsive energy release
\textbf{or}
      very small (EUV) brightening& $\sim 10^{24}$&?$\ll$A&?$\ll 10^{-8}$ &  ?
\\
  \hline\noalign{\smallskip}
\end{tabular}
\end{centering}
\newline $^1$ ``Active Region Transient Brightening'' observed in SXRs \citep{shimizu1995}
\end{table}
\index{flares!classification!table}\index{flare classification!table}
\index{ARTB}

Flare phenomena encompass all of the accessible wavelength ranges of
electromagnetic radiation, as well as the emission of neutral particles, gaseous
ejecta, and large-scale shock waves \citep[e.g.][]{chapter2}. 
In the most energetic flares, all of these phenomena appear, considerably more intensely
than in smaller flares and regardless of the detailed physics; this is the so-called
\emph{big-flare syndrome} \citep[e.g.,][]{1982JGR....87.3439K}.
\index{syndromes!big-flare}
\index{big-flare syndrome}
This makes these events helpful diagnostically. This property suggests that if a particular
phenomenon (e.g., white-light continuum) is not observed in a weaker flare (as
not distinguishable from the background), the process may still function in the
same way\index{white-light flares!big-flare syndrome}. 
Non-scalable properties might exist\index{caveats!non-scalable properties}; 
one that is often cited is the
maximum energy attained by accelerated particles.
An emission such as $\pi^0$-decay $\gamma$-ray emission might thus have a non-linear threshold
dependence; see \citet{chapter4}.
\index{gamma-rays!pi@$\pi^0$-decay in microflares}
Of course, given that there weak flares are considerably more numerous, exceptional cases can often be found, such as microflares with nonthermal emission to remarkably high energies
\citep[e.g.,][]{2008A&A...481L..45H}.

\subsubsection{Flare relationship to CMEs and SEPs}\label{sec:hannah_cme}

\begin{figure}\sidecaption\centering
\includegraphics[width=0.5\columnwidth]{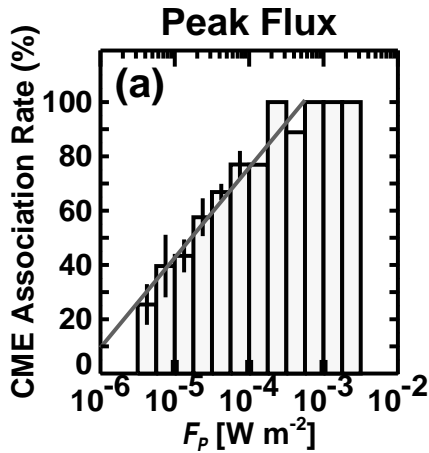}
\caption{\label{fig:hannah_cme}The association rate of CMEs with flares as
a function of the flare peak \textit{GOES} flux \citep{2006ApJ...650L.143Y}. Only
flares from C-class ($10^{-6}$~W~m$^{-2}$) to X-class  ($>$10$^{-4}$~W~m$^{-2}$) are
included in this survey. Reproduced by permission of the AAS.}
\end{figure}

There has been a long debate whether ejection of material from the corona,
namely a Coronal Mass Ejection (CME), triggers a flare, or vice-versa, but it is now
generally accepted that they are both consequences of coronal energy release
through the reconfiguration of the magnetic field.
\index{microflares!and CMEs}
\index{coronal mass ejections (CMEs)}
\index{coronal mass ejections (CMEs)!and microflares}
\index{solar energetic particles (SEPs)!and microflares}
The issue of whether a flare or CME is produced will depend on the magnetic field configuration
\citep{2007ApJ...665.1428W}. Using a large survey of nearly 7000 CMEs,
\citet{2004JGRA..10907105Y} found that all the most energetic ($>$X2) flares have
CMEs; see Figure~\ref{fig:hannah_cme} \citep{2006ApJ...650L.143Y}. The largest
and fastest CMEs are generally associated with such large flares. The rate for
C-class flares is about 20\% and for M-class about 40\%. Of course, there may be a
sensitivity issue which limits the ability to observe the small CMEs associated with
microflares. Studies of flares that do not have associated CMEs have found these
events to have a correlation between peak intensity and flare duration
\citep{2003AdSpR..32.1051K}. No such correlation was found for CME flare events,
suggesting a physical difference in the time development of a flare when a CME is
involved. CMEs appear to have broadly the same range of characteristics whether
or not they are associated with a flare, but the fastest and broadest CMEs are
always associated with the most energetic flares \citep{2005A&A...435.1149V}. In
terms of the energy partition between flares and CMEs, the CME kinetic energy
ranges may be comparable to the total flare energy \citep{2005JGRA..11011103E}.

\index{flares!association with CMEs!illustration}

The ejection of highly accelerated (up to several MeV) electrons, protons and
heavy ions from the Sun into interplanetary space, called Solar Energetic Particles
(SEPs), are sometimes associated with flares \citep[e.g.][]{1999SSRv...90..413R}\index{solar energetic particles (SEPs)}.
Impulsive SEP events\index{solar energetic particles (SEPs)!impulsive events} 
demonstrate a rapid burst of electron enhancement often
related to the impulsive HXR phase of flares, suggesting that the same coronal energy
release is responsible for the flare and SEPs. These energetic electrons therefore
provide an important diagnostic of flare particle acceleration, especially when
combined with HXR flare observations\index{radio emission!type III burst}.
Type III radio bursts are also produced by energetic electrons escaping from the Sun into interplanetary space, see Section~\ref{sec:hannah_mfrad}. 
Gradual SEP events\index{solar energetic particles (SEPs)!gradual events} show a slower enhancement of energetic protons, thought to be accelerated by a CME-associated shock\index{shocks!interplanetary} in
interplanetary space rather than in the coronal energy release. 
Although these proton events are not directly related to flares, \citet{1995ApJ...453..973K} showed
that flares that were associated with 10~MeV proton events predominantly
demonstrated progressive hardening of the HXR photon spectra into the decay
phase of the HXR emission. 
In contrast, the other population of HXR events not
associated with SEPs exhibit softening spectra in the decay phase. A similar result
was found for X-class flares during the January 2005 solar storm event
\citep{2008ApJ...673.1169S}. 
\index{coronal mass ejections (CMEs)!and soft-hard-harder pattern}
In these flares, four out five showed spectral
hardening with \textit{RHESSI} and were associated with interplanetary proton events. A
study of 37 events with \textit{RHESSI} found that the majority of events that
demonstrated HXR spectral flattening (12 out of 18) produced SEPs (detected with
\textit{GOES} $>$10~MeV proton data and \textit{WIND}/3DP $0.1-1$~MeV proton and $30-500$~keV
electron data) and all without flattening (19 flares) did not produce SEPs
\citep{2009ApJ...707.1588G}.
\index{satellites!WIND@\textit{WIND}!3DP}

\subsection{Microflares}\label{sec:hannah_micro}
\index{microflares}

\subsubsection{Association with active regions}\label{sec:hannah_microar}
\index{flares!H$\alpha$}\index{flare classification!H$\alpha$}
\index{active regions!association with microflares}

``Subflares'' had always been known to the H$\alpha$ observers
\citep[e.g.,][]{1963QB528.S6.......} as events that were small in area but not
necessarily faint. 
The term ``microflare'' was introduced in the 1980s by
\cite{1983SoPh...89..287S} for SXRs, and by \cite{1984ApJ...283..421L} for HXRs.  The
presence of X-rays, the occurrence of subflares with ``brilliant'' H$\alpha$
classification (high intensity), and the discovery of faint microwave bursts
\citep{1994ApJ...437..522G,1997ApJ...477..958G} in association with SXR
microflares made it clear that the basic flare physics extended over a wide
magnitude range.
Flares generally (including subflares and microflares) show both thermal and
nonthermal emission, indicating the presence of particle acceleration and plasma
heating. 
\index{satellites!GOES@\textit{GOES}}
In terms of \textit{GOES} classification these events are typically A- and B-class
events, down to and beyond the sensitivity limit of the \textit{GOES} detectors.

\begin{figure}\centering
\includegraphics[height=50mm]{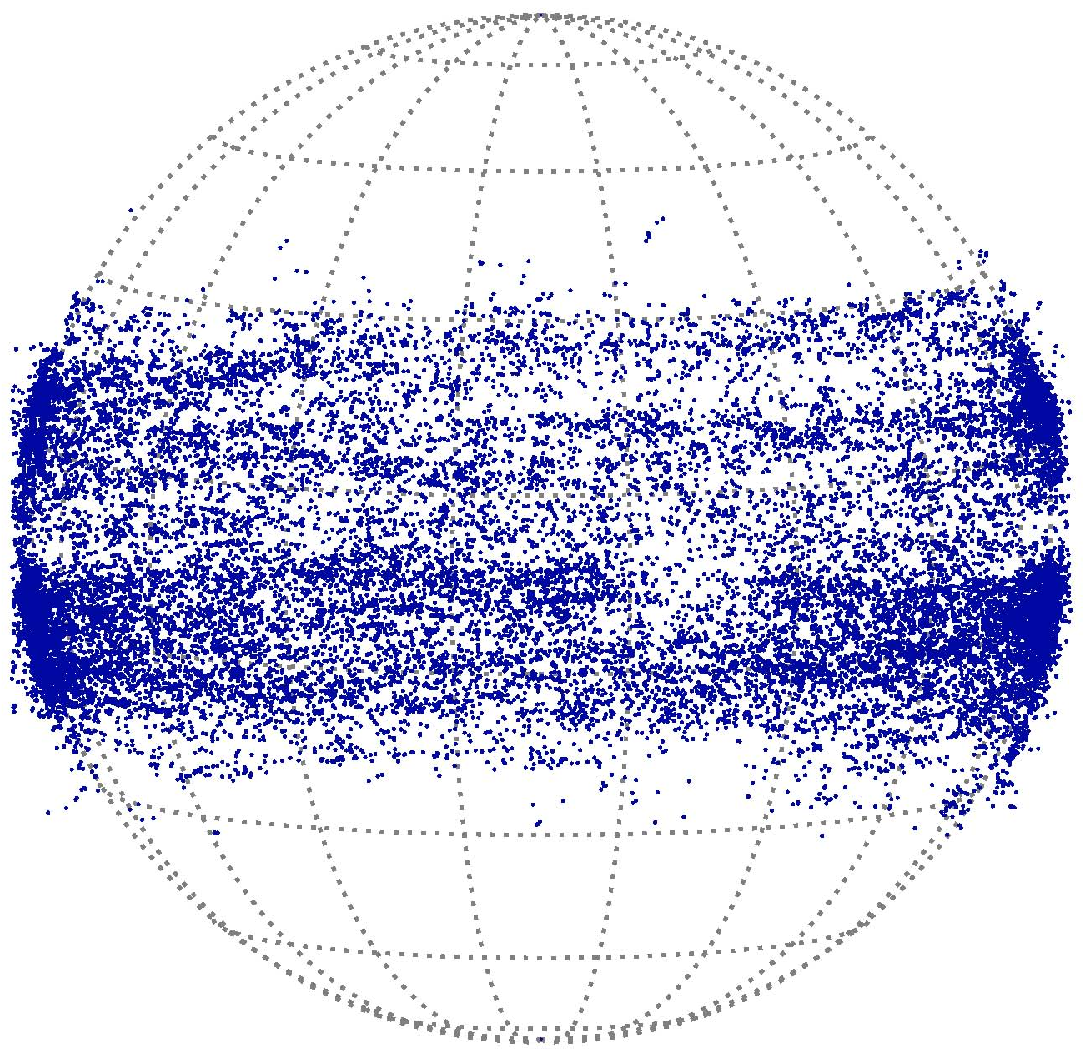}
\includegraphics[height=40mm]{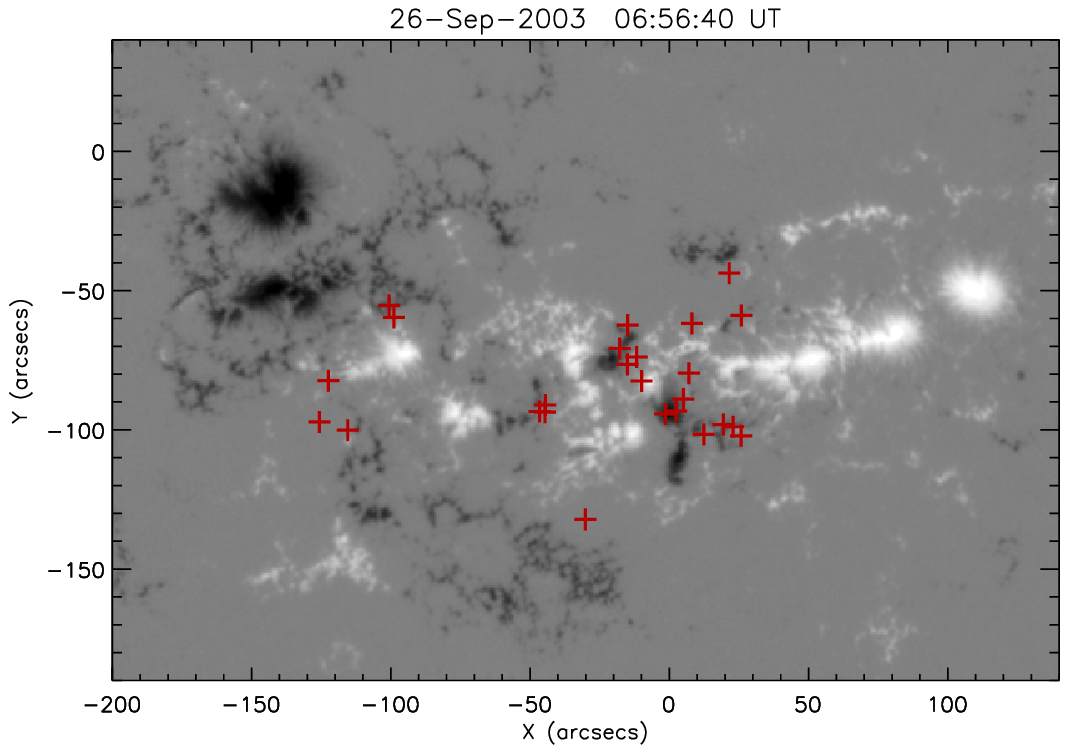}
\caption{\label{fig:hannah_mfpos} (\textit{Left}) The locations of 24,097
microflares observed with \textit{RHESSI} between March 2002 and March 2007
\citep{christe2008}. All of the confirmed events could be associated with
active regions. Reproduced by permission of the AAS. (\textit{Right})
The locations of one day of \textit{RHESSI} microflares
from active region 10456 \citep{2007SoPh..246..339S}. Of 53 events found, 24
could be imaged. The centroid positions (3-6 keV) are indicated by crosses and
overplotted on the MDI (Michelson Doppler Imager of \textit{SOHO}) magnetogram.}
\end{figure}
\index{active regions!association with microflares!illustration}

The HXR observations prior to \textit{RHESSI} normally suffered from poor sensitivity due
to lack of access to lower hard X-ray energies ($<$15~keV), hence these
observations missed most of the microflares. The problem was the use of thick
fixed attenuators to reduce the excessive low energy counts in large flares. One
exception to this was the HXIS\footnote{Hard X-ray Imaging Spectrometer.}
imager on the \textit{Solar Maximum Mission (SMM)}
\index{HXIS}
\citep{1983SoPh...89..287S}, which covered the $3-30$~keV range with a restricted
field of view and very small detectors, thus avoiding saturation. Because of this,
the main HXIS results were in a spectral region comparable with the \textit{GOES} data. 
At lower energies ($0.25-4$~keV), the Soft X-ray Telescope (SXT) on \textit{Yohkoh} found
microflares to occur in active regions \citep{shimizu1995}.
\index{satellites!Yohkoh@\textit{Yohkoh}}
\index{Yohkoh@\textit{Yohkoh}!Soft X-ray Telescope (SXT)}
\index{satellites!SMM@\textit{SMM}!HXIS}.

The first observation of HXR microflares was with balloon-borne detectors that
could observe only down to about 15~keV due to atmospheric absorption\index{absorption!Earth's atmosphere}, and
which had high sensitivity from the use of large-area, low-background detectors
\citep{1984ApJ...283..421L}. 
\index{satellites!Yohkoh@\textit{Yohkoh}!HXT}
Subsequently with \textit{Yohkoh}/HXT, emission at
$14-23$~keV was found to be associated with SXR brightenings in active regions
\citep{1997ApJ...491..402N}.
Microflares were also detected at $8-13$~keV with
\textit{CGRO}/BATSE \citep{2001ApJ...557L.125L}\index{satellites!CGRO@\textit{CGRO}}.
However, it was not until \textit{RHESSI} that the
detailed associations of these HXR microflares could be readily investigated.

\textit{RHESSI}'s introduction of a movable attenuating shutter 
system\index{RHESSI@\textit{RHESSI}!attenuating shutters}
\citep{2002SoPh..210....3L} allowed the attenuators only to be deployed when the
detectors were saturated. With no shutters, the detectors become saturated once
solar emission (either background emission from active regions or flares) is
approximately greater than \textit{GOES} C1 level. \textit{RHESSI} microflares are therefore flares
that occur when the attenuating shutters are out; they are typically sub-C-class events, 
$<$10$^{-6}$W~m$^{-2}$ in the $1-8$~\AA~\textit{GOES} band. 
The full view and sensitivity of \textit{RHESSI} is
available to such events, providing the first comprehensive HXR view of
microflares, with both imaging and spectroscopy. In addition to this, \textit{RHESSI}
introduced~Ge detectors with resolution much superior to the scintillation
counters previously used for~HXR observations, allowing the properties of steep
spectra to be accurately determined.

\begin{figure}\centering
\includegraphics[height=48mm]{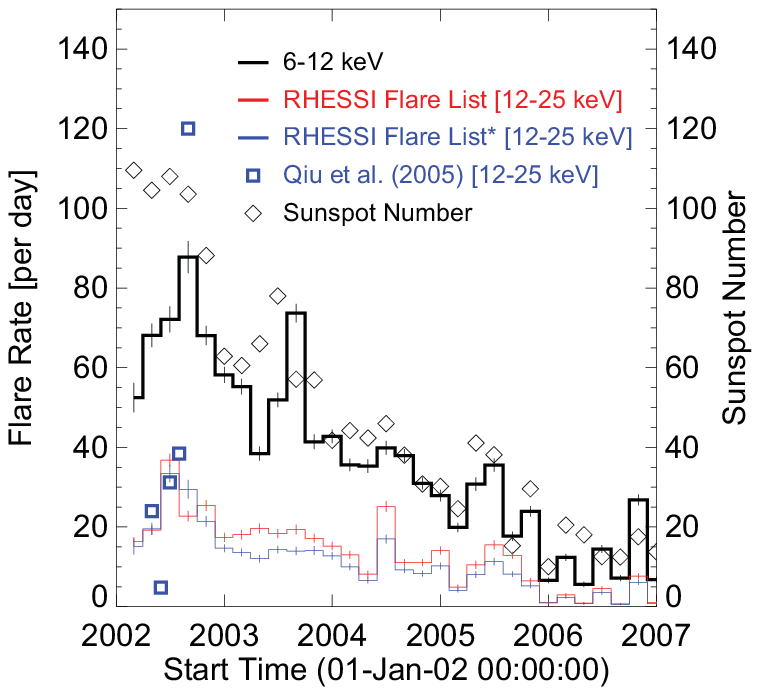}
\includegraphics[height=50mm]{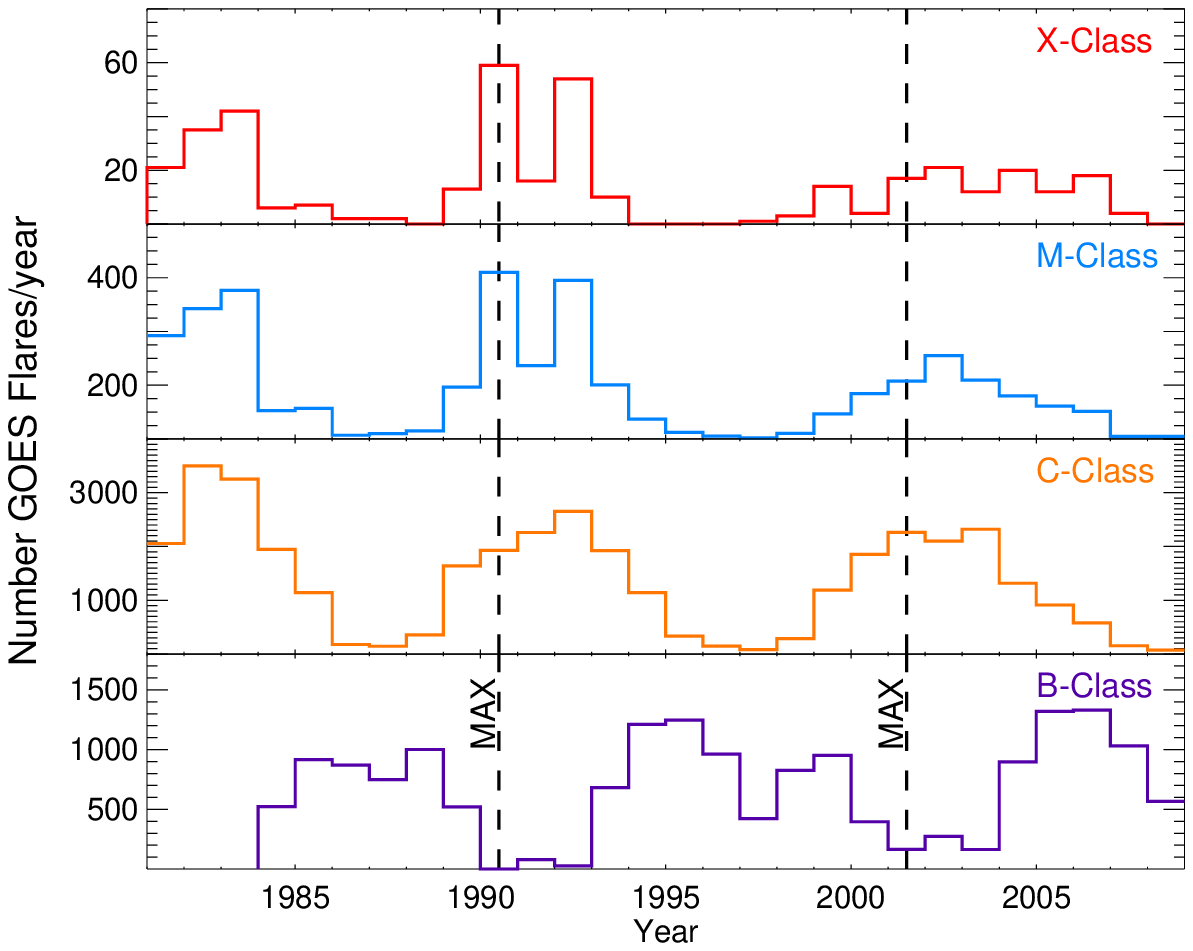}
\caption{
\label{fig:hannah_mftime}(\textit{Left}) Occurrence rate of \textit{RHESSI} microflares
($<$C1 events) \citep{christe2008}, showing a clear decline over the waning
years of Cycle~23.  Reproduced by permission of the AAS. (\textit{Right}) The
number of \textit{GOES} flares per year per flare class for X-class (\emph{top}) through
M, C to B-class (\emph{bottom}). The microflares seem to be
anti-correlated with the larger flares, which would be expected from the increased
background during times of high activity. 
Note that B-class flare data was only available from 1984 \citep[data from][]{milliganp}.}
\end{figure}

\textit{RHESSI} microflares appear to be exclusively localized to active regions.\index{microflares!active-region association}\index{quiet Sun!absence of microflares}
Figure~\ref{fig:hannah_mfpos} (left panel) shows this explicitly with the positions
of 24,097 events, all of which are identified as coming from active regions
\citep{christe2008}. 
These events were identified from analyzing all count-rate spikes in the
\textit{RHESSI} $6-12$~keV energy range between March 2002 and March 2007, at times when the
attenuating shutters were out.
\index{RHESSI@\textit{RHESSI}!attenuating shutters!and microflares}
\index{imaging!microflares}\index{microflares!imaging}
A total of 25,705 events were found, though only 24,097 events
could be imaged to give positions on the solar disk. The events not counted
include some of the faintest true microflares, but are predominantly misidentified particle events
(which gave a non-solar disk position) or events close to the rotation
axis of \textit{RHESSI} (events close to where the spacecraft rotation axis is
pointing will have no modulation in their detected signal and hence no spatial
information can be recovered).
\index{caveats!RHESSI@\textit{RHESSI} rotation axis}\index{RHESSI@\textit{RHESSI}!rotation axis and imaging}
There is repeated microflaring from active regions, thus
tracing out the location of a region as it moves across the disk.
The microflaring rate varies greatly between active regions. 
Detailed \textit{RHESSI} imaging
of all the microflares from one active region compared to MDI magnetograms
shows that the microflare emission occurred at multiple locations throughout the
active region \citep{2007SoPh..246..339S}, though many originated in one localized
area of the region (Figure~\ref{fig:hannah_mfpos}, right panel). Another sample
of microflares showed that they mostly occur near magnetic neutral lines in active
regions \citep{2004ApJ...604..442L}\index{microflares!association with magnetic neutral line}.

As expected, then, Figure~\ref{fig:hannah_mftime} (left panel) shows that these
events tend to follow the solar cycle in their occurrence frequency\index{microflares!solar-cycle dependence}.
We note that it
is difficult to follow the microflare rate over a whole solar cycle because the solar
X-ray background level also follows the solar cycle. Fewer microflares are
observed during periods of high solar activity as they are hidden by the high
background from active regions.\index{active regions!as source of background radiation}
This can be clearly seen in the number of \textit{GOES} B-class flares (right panel of 
Figure~\ref{fig:hannah_mftime}).
\index{microflares!observational bias towards low activity} 
Their rate appears unexpectedly to be almost anti-correlated with the solar cycle. 
This observational bias means that the microflaring rate is only accurately known during times of
lower activity/solar background.

As the HXR microflares observed by \textit{RHESSI} are inherently  active-region
phenomena, it seems entirely reasonable to associate them with microflares
\citep{1983SoPh...89..287S} and the active-region transient brightenings (ARTBs)
seen in soft X-rays \citep{shimizu1995}.
\index{active-region transient brightenings!identification with microflares}
This had been previously postulated
\citep{1993pssc.symp..113T} and early evidence suggested it to be correct
\citep{1997ApJ...491..402N}, but the confirmation was only possible with \textit{RHESSI's}
hard X-ray sensitivity.

\subsubsection{Physical properties}

\begin{figure}\centering
\includegraphics[height=65mm]{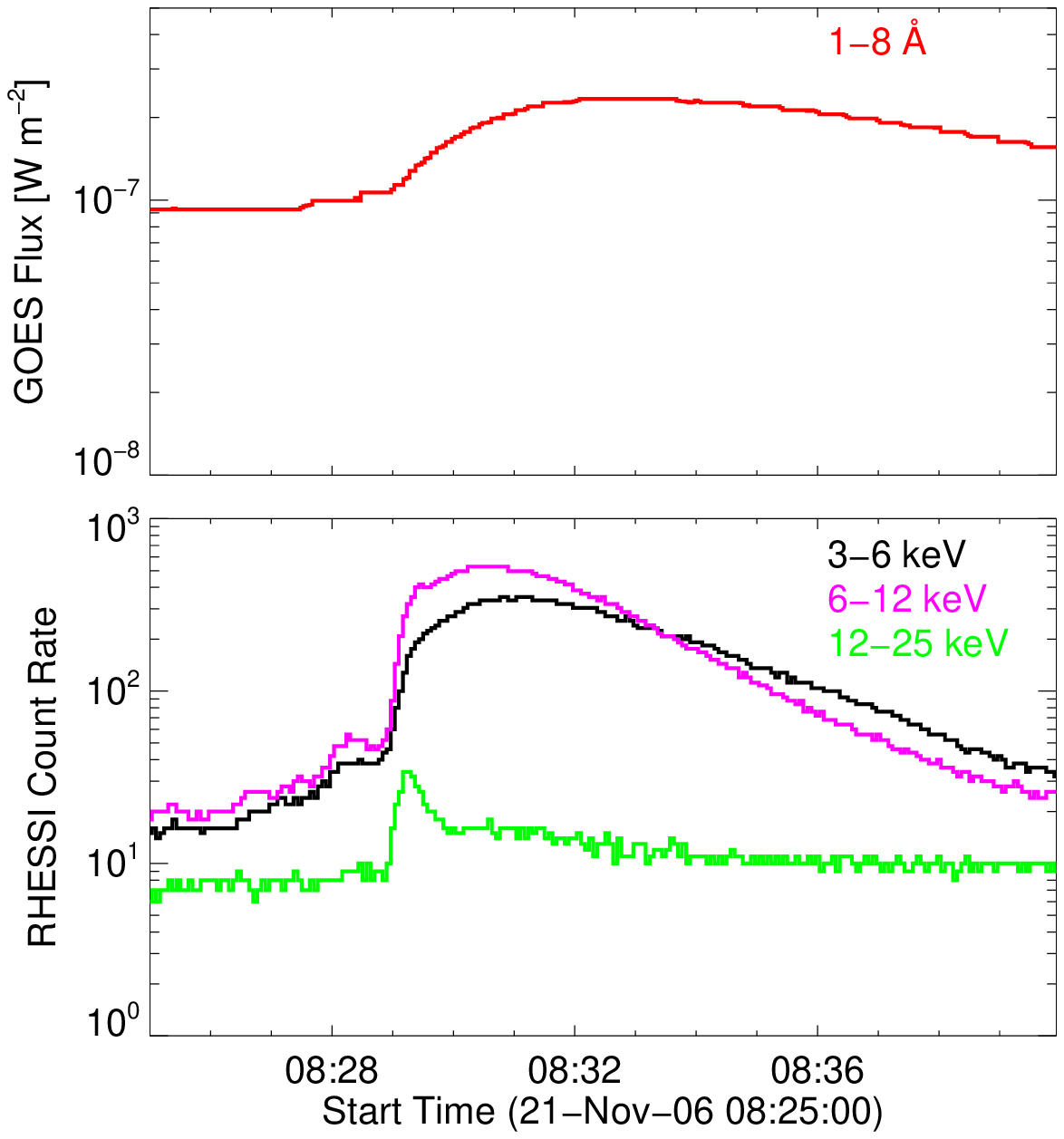}
\includegraphics[height=65mm]{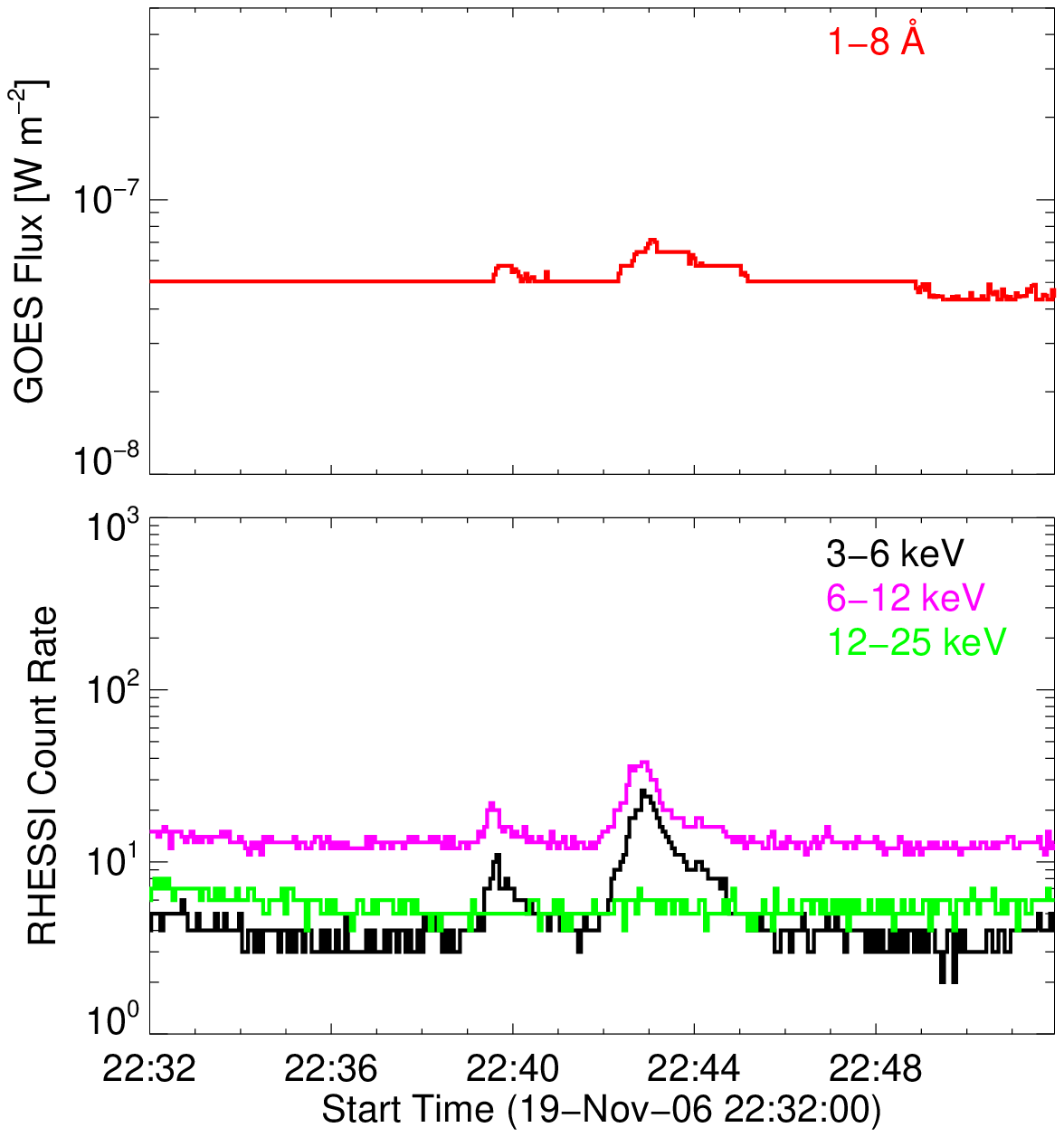}
\caption{
\label{fig:hannah_mfltc} The \textit{GOES} $1-8$~\AA~lightcurve (top panels) and \textit{RHESSI}
lightcurves for some example microflares. (\textit{Left:}) A single B-class
microflare (B1 with pre-flare background subtracted) which shows the classic
flare time profile: impulsive at higher energies with a slower rise and more
gradual fall at lower energies. This event also shows some pre-flare emission.
(\textit{Right:}) two A-class events (A1 and A3, background-subtracted) both
of which have no emission above 12~keV observed by \textit{RHESSI}; their nonthermal
emission hidden by the background.
Note that the background level in these microflares is predominantly solar for \textit{GOES},
but terrestrial and instrumental in origin for \textit{RHESSI}.}
\end{figure}
\index{flare (individual)!SOL2006-11-06T08:32 (B1)!illustration}
\index{flare (individual)!SOL2006-11-06T22:40 (A1)!illustration}
\index{flare (individual)!SOL2006-11-06T22:43 (A3)!illustration}

The X-ray time profiles of typical microflares (examples given in Figure
\ref{fig:hannah_mfltc}) demonstrate a similar structure to those in typical large
flares: a short impulsive burst at higher energies (HXRs) followed by a more gradual
thermal emission at lower energies (SXRs). Microflares also demonstrate other
temporal phenomena seen in large flares like more gradual emission or
pre-impulsive behavior though the timescales are generally shorter (seconds to
minutes instead of several to tens of minutes).

\begin{figure}
\centering
\includegraphics[height=50mm]{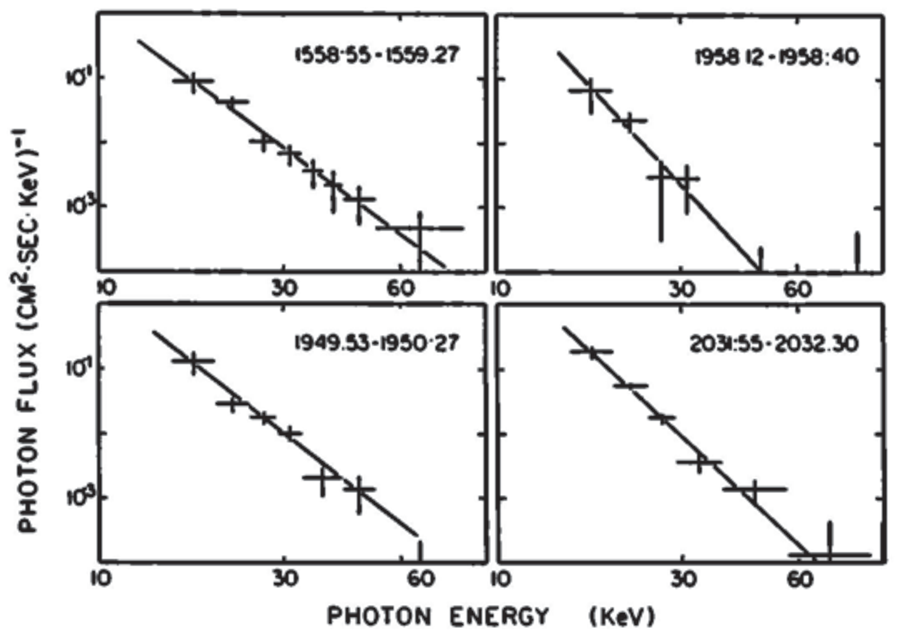}\\
\includegraphics[height=40mm]{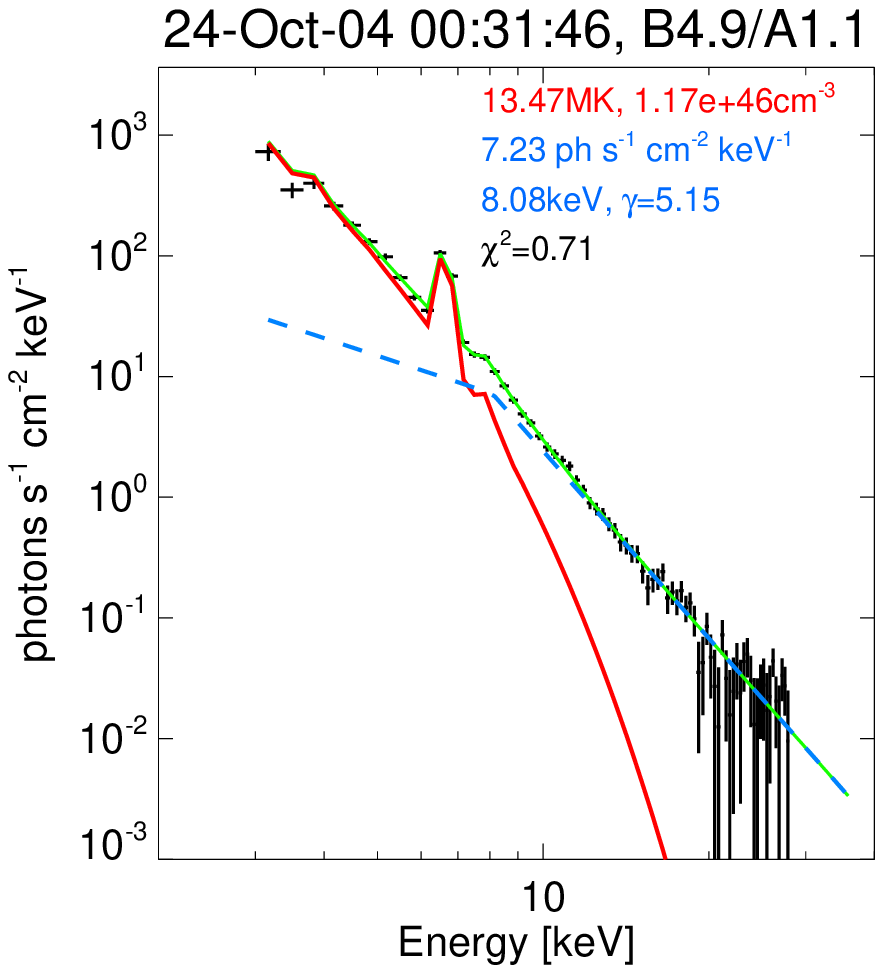}
\includegraphics[height=40mm]{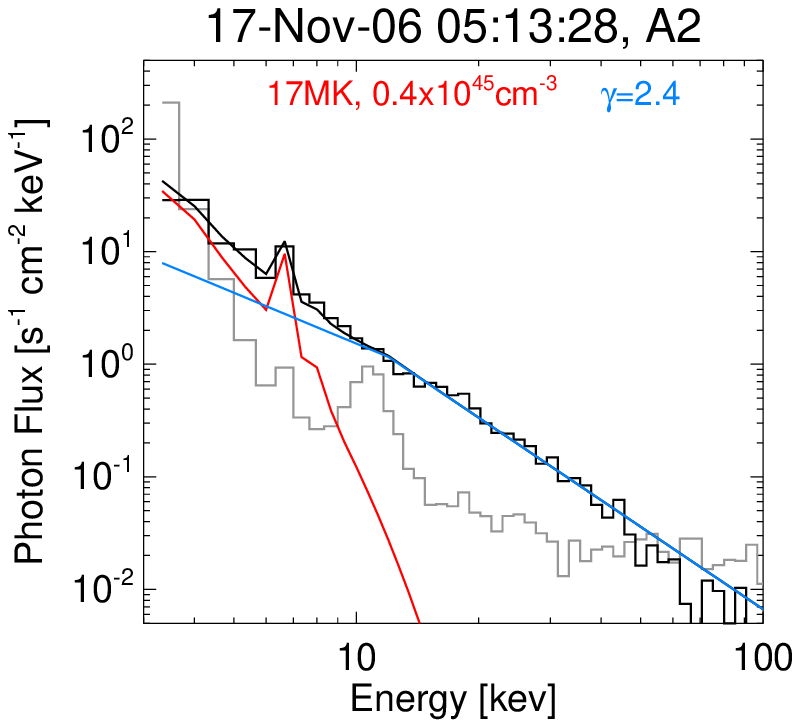}
\caption{
\label{fig:hannah_mfspec} (\emph{Top four panels:}) the first HXR microflare spectra
from a balloon-borne instrument \citep{1984ApJ...283..421L}, and (\textit{bottom two panels:} spectra from
\textit{RHESSI} showing a typical microflare \citep{hannah2008} (\emph{left})
and one with a particularly hard spectrum (\emph{right})
\citep{2008A&A...481L..45H}.
The red and blue lines are the thermal and
nonthermal components of the model, respectively. 
In the \textit{RHESSI} spectra, the
upper limit of the nonthermal component is hidden by the instrumental
background (the large errors bars in the bottom-left spectrum and the grey
histogram in the bottom-right spectrum).
Reproduced by permission of the AAS.}
\end{figure}

The first HXR spectra of microflares (top panel of Figure~\ref{fig:hannah_mfspec})
showed only a power-law component, characteristic of nonthermal emission, due
to the detector being sensitive only above about 13 keV
\citep{1984ApJ...283..421L}. \textit{RHESSI} spectra (bottom row of Figure
\ref{fig:hannah_mfspec}) can be obtained with moderate energy resolution down
to a few keV. This allows the thermal component, and the transition from it to the
nonthermal component, to be seen. These spectra (from the time of peak
emission just after the impulsive phase) are similar to those seen in more
energetic flares, except that (a) the thermal emission tends to be at lower
temperatures and (b) the nonthermal component tends to have a steeper
spectrum (fewer electrons accelerated to high energies). In the bottom left panel
of Figure~\ref{fig:hannah_mfspec}, we have a typical \textit{RHESSI} microflare with a
fitted temperature of about 13~MK.
The Fe~feature\index{Fe lines!in microflare spectra}\index{microflares!Fe line}
\citep[e.g.,][]{2004ApJ...605..921P} is clearly in evidence at about 6-7~keV and a
nonthermal spectrum above 8~keV with an E$^{-5.15}$ power law.
\index{hard X-rays!microflares}\index{microflares!non-thermal HXR} 
The other \textit{RHESSI}
spectrum shown (bottom right panel of Figure \ref{fig:hannah_mfspec}) is of an
anomalously hard microflare, which despite being a \textit{GOES}~A2 event (near the limit
of the \textit{GOES} sensitivity) demonstrates strong nonthermal emission to relatively
high energies, just as in much more powerful events \citep{2008A&A...481L..45H}.
One should note, however, that the early microflare HXR spectra of
\cite{1984ApJ...283..421L} were more sensitive above about 30 keV since their
shielded detectors were able to reduce the background clearly evident in the
\textit{RHESSI} spectra\index{microflares!steep hard X-ray spectra}.
These Lin et al. (1984) microflares also demonstrate steep nonthermal emission of
$I(\epsilon) \propto \epsilon^{-\gamma}$ with $\gamma=4$ to $6$ that continue
to energies normally obscured by background in \textit{RHESSI} observations.

\begin{figure}\centering\sidecaption
\includegraphics[width=0.65\columnwidth]{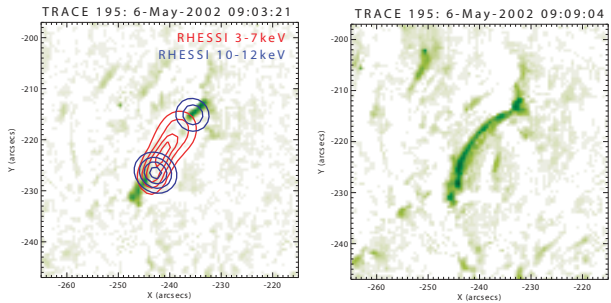}
\caption{\label{fig:hannah_mfimage}
Overlay of \textit{RHESSI} thermal (red) and nonthermal (blue) contours on \textit{TRACE},
showing the typical flare and microflare scenario of (\textit{left}) footpoints
depositing their energy to heat chromospheric material that expands upwards,
filling the loops connecting them. (\textit{Right:}) his material then cools, emitting
brightly in EUV. Updated version of figure
from \citet{2002SoPh..210..445K}.}
\end{figure}
\index{flare (individual)!SOL2002-05-06T09:01 (A8.0)!illustration}
\index{footpoints!microflares!illustration}

\begin{figure}\centering\sidecaption
\includegraphics[width=0.5\columnwidth]{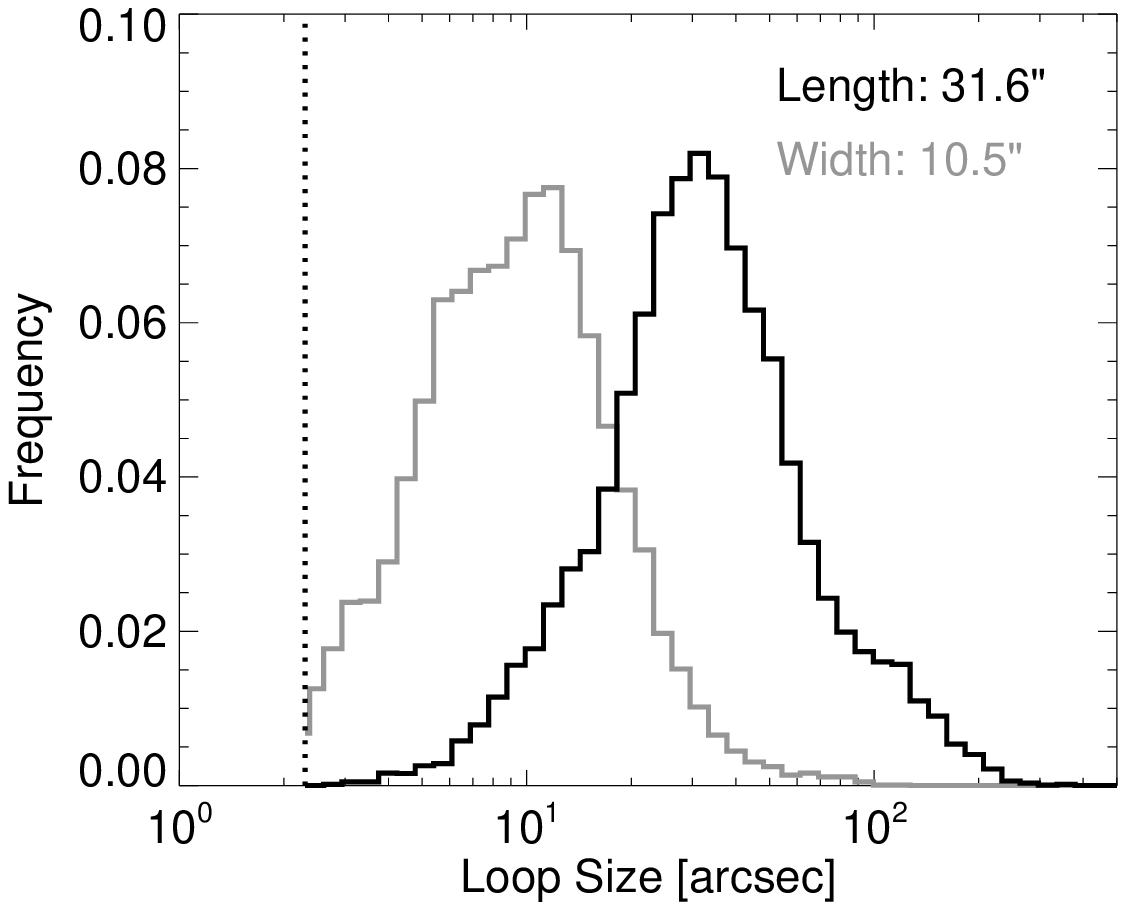}
\caption{\label{fig:hannah_mflength}
Loop arc length and width distributions for 18,656 microflares imaged by \textit{RHESSI}
indicating that they are predominantly elongated structures \citep{hannah2008}.
Thus microflares are not necessarily spatially small. Reproduced by permission
of the AAS.}
\index{footpoints!microflares!illustration}
\end{figure}

\textit{RHESSI} microflares are not spatially small and typically show an elongated
loop-like structure, with higher-energy HXR emission from the footpoints at the
ends of the loop\index{footpoints!microflares!HXR}
\citep[e.g.,][]{2002SoPh..210..445K,2004ApJ...604..442L,2007SoPh..246..339S,
2008A&A...481L..45H}. 
Figure \ref{fig:hannah_mfimage} shows an example of this
overlaid on an EUV image from \textit{TRACE}.
\index{magnetic structures!microflare image morphology}
We interpret the EUV loop as the cooling post-flare loop
\index{loops!in microflares} 
which was first seen at higher temperatures in the \textit{RHESSI} image.
This hot material could be driven (evaporated) from the chromosphere\index{chromospheric evaporation} by the energy deposition of the nonthermal electrons penetrating to the loop footpoints
\citep{fisher1985,abbett1999,brosius2004} and producing HXR sources there.
Several \textit{RHESSI} microflares have also been observed with the HXR loop connecting
two bright H$\alpha$ kernels \citep{2004ApJ...604..442L}, another possible
signature of the energy deposition of nonthermal electrons in the lower
atmosphere\index{loops!H$\alpha$ footpoints in microflares}\index{footpoints!microflares}. 
Previous work had shown the association between microflare SXR
and H$\alpha$ emissions \citep{2002ApJ...574.1074S}, consistent with the
relationship known from ordinary flares \citep{1971SoPh...16..431T}.

Figure~\ref{fig:hannah_mflength} shows the fitted lengths and widths of the
thermal emission (4-8 keV) for 16 seconds about the time of peak emission in 6-12
keV for almost 19,000 \textit{RHESSI} microflares \citep{hannah2008}. These parameters
are from simple forward-fitted image models\index{inverse problem!microflare imaging} and show that the microflares have
a clear tendency to be elongated, as we would expect from their identification
with the loops that characterize SXR microflares \citep[e.g.,][]{shimizu1995}.
\index{loops!microflare!lengths}
The distribution of the loop length has a peak well above \textit{RHESSI}'s angular resolution,
so we conclude that these microflares may not be inherently small-scale. Many
ordinary flares, though more energetic, have dimensions in this range. In addition,
these lengths are not correlated with the microflare peak \textit{GOES} flux or peak
\textit{RHESSI} emission \citep{hannah2008}.

\subsubsection{Associated ejecta and radio emission}\label{sec:hannah_mfrad}
\index{microflares!radio emission}

Radio signatures generally guide us to the corona, since the opacity (and the
minimum height of formation of the emission) increases in height almost
monotonically with wavelength \cite[e.g.,][]{chapter5}. Millimeter-centimeter
wavelengths (microwave GHz emission) originate mainly in the lower
atmosphere, meter-waves (100s MHz to MHz radio emission) in the lower corona,
and the longest wavelengths right out into the solar wind. The longer-wavelength
radio observations thus often show ejecta, especially via nonthermal plasma
signatures \cite[e.g.,][]{2008A&ARv.tmp....6P}.

The microflare gyrosynchrotron emission\index{radio emission!gyrosynchrotron!microflares} is closely correlated in time to the HXR emission \citep{2004ApJ...612..530Q,2006A&A...451..691K}, indicating
that they are both signatures of the accelerated flare electrons. The previous
observations of SXR microflares also demonstrated a correlation with microwave
emission \citep{1997ApJ...477..958G,1999ApJ...513..983N}. 
The microwave spectral index for microflares is found to be flatter (harder) than the \textit{RHESSI} HXR spectra \citep{2004ApJ...612..530Q}. 
This is thought to be due to the microwave emission
originating from electrons with higher energies than those producing X-rays
\citep{1986SoPh..105...73N}. 
This discrepancy in spectral indices is also observed in larger flares \citep{2000ApJ...545.1116S} though it is possibly greater in microflares. 
This may be due to the HXR spectra being typically steeper in microflares than large flares. 
A B-class microflare was observed to have nonthermal microwave emission from MeV-energy electrons but showed no HXRs at lower energies \citep{1999ApJ...522..547R}. 
This behavior could result from trapping of the lower-energy electrons during the peak of the microwave emission from the higher-energy electrons.
\index{footpoints!microflares!microwave emission}
\index{microflares!microwave emission}
\index{microflares!and MeV electrons}
\index{microflares!gyrosynchrotron emission}
\index{gyrosynchrotron emission!microflares}
\index{electrons!accelerated!in microflares}
\index{trapping!and microwaves}

\begin{figure}
\centering
\includegraphics[height=70mm]{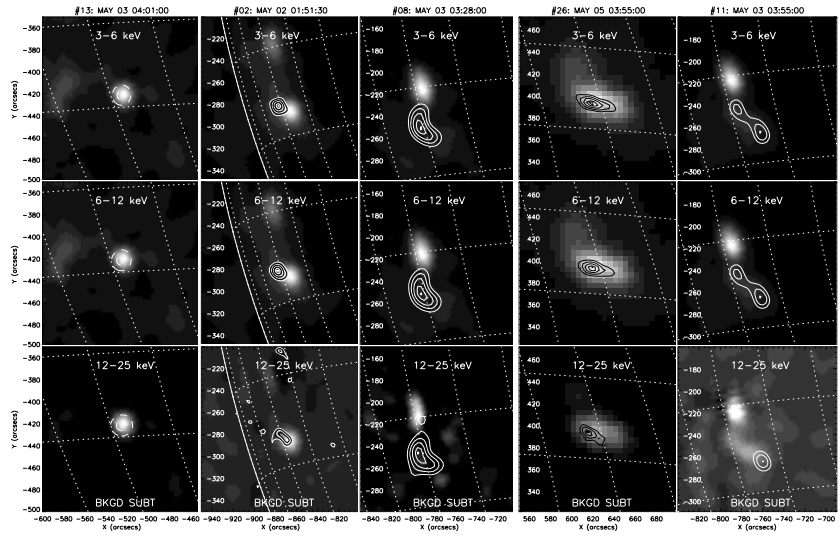}
\caption{\label{fig:kundu_HXRmicro}
Five microflares (one per column) imaged in
different HXR energy bands (increasing from top row to bottom) with \textit{RHESSI}
(contours), overplotted on NoRH 17~GHz images \citep{2006A&A...451..691K}. The
bottom microwave image is background-subtracted. Similar-sized footpoints and
loop structures are seen in both HXRs and microwaves.}
\index{microflares!illustration}
\end{figure}

Imaged microwave emission using the Nobeyama radioheliograph (NoRH) for
\textit{RHESSI} microflares indicated similarly sized and separated footpoints in both HXR
and microwave emission \citep{2006A&A...451..691K}; see Figure
\ref{fig:kundu_HXRmicro}. The higher-frequency microwaves come from the
footpoints, and the lower-frequency from the connecting loop
\citep{2005AdSpR..35.1778K,2006A&A...451..691K}.\index{Nobeyama radioheliograph}\index{observatories!NoRH}

\begin{figure}
\centering\sidecaption
\includegraphics[height=90mm]{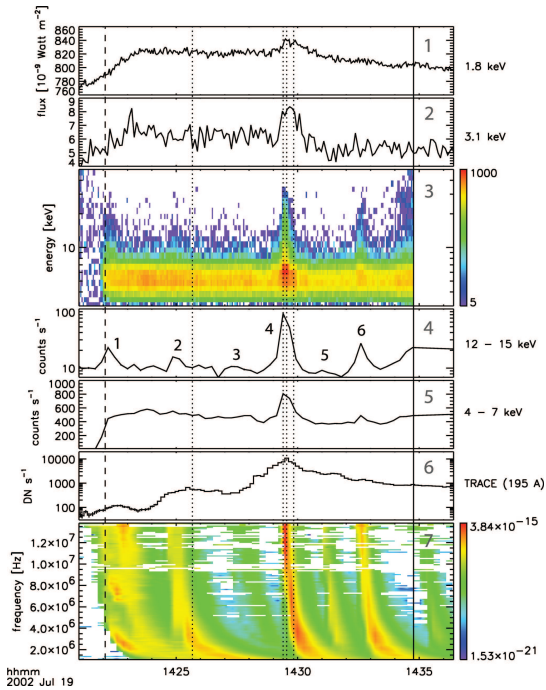}
\caption{\label{fig:christe_t3}The lightcurves and radio spectrograms of six \textit{RHESSI}
microflares from  July 19, 2002 \citep{2008ApJ...680L.149C}. 
The top two panels
show the \textit{GOES} lightcurves in $1-8$~\AA~(with a B8~background level) and
$0.5-4$~\AA. The next three panels show the \textit{RHESSI} spectrogram with night-time
background removed and lightcurves in $12-15$~keV (nonthermal) and $4-7$~keV
(thermal). The EUV lightcurve from \textit{TRACE} and a radio spectrogram
from \textit{WIND}/WAVES are shown in the bottom two panels. 
The radio emission starts at around 14~MHz and drifts
down to 2~MHz. It is well correlated with the \textit{RHESSI} HXR lightcurve.
Reproduced by permission of the AAS.}
\index{satellites!WIND@\textit{WIND}!WAVES}\index{microflares!illustration}\index{radio emission!type III burst!illustration}
\end{figure}

\textit{RHESSI} microflares are often associated with Type~III radio bursts\index{microflares!and type III bursts},  produced by
electron beams escaping from the corona and often subsequently detected as
SEPs\index{radio emission!type III burst}.
These are the ``fast drift'' bursts identified with weakly relativistic electron
beams \cite[e.g.,][]{1963ARA&A...1..291W}\index{hard X-rays!and with type III bursts}.
The HXR emission in these cases still
seems to be a signature of accelerated electrons reaching the lower atmosphere
at loop footpoints, although the radio emission comes from accelerated electrons
of similar energies moving outwards from a postulated coronal acceleration site.
An example of six microflares identified with \textit{RHESSI} and their associated MHz
radio emission observed with \textit{WIND}/WAVES is shown in Figure \ref{fig:christe_t3}.
Here we can see the time correlation of the HXR emission and the high-frequency
radio emission. 
The microflare with the largest HXR flux produced the brightest radio burst. 
The association with Type~III bursts establishes that electrons
accelerated in the microflares have access to open field lines (or field lines that
extend into the upper corona), not requiring an eruption prior to the event.
\index{microflares!and open field lines}
An association between Type~III bursts and small SXR flares had previously been
found by \cite{1982A&A...107..178F}.

The presence of open field lines in microflares has also been seen with the
association of jets\index{microflares!jet association} of material flowing out of the flare region
\citep{1996PASJ...48..123S}. For \textit{RHESSI} HXR microflares, such jets have been
observed in EUV with \textit{TRACE} \citep{2008ApJ...680L.149C} and in EUV and SXRs with
\textit{Hinode} \citep{2008A&A...491..279C}\index{satellites!TRACE@\textit{TRACE}}.
\index{satellites!Hinode@\textit{Hinode}}
In the former case, multiple microflares and
\index{radio emission!type III burst!EUV jets}
type~III burts were observed, occurring every few minutes, with EUV jets associated
with the largest microflares \citep{2008ApJ...680L.149C}.\index{jets!recurring}\index{jets!X-ray}\index{reconnection} 
In the latter case, the
recurring jets occurred on a timescale of hours and were attributed to
chromospheric evaporation flows due to recurring coronal magnetic reconnection
\citep{2008A&A...491..279C}.

This association with jets is also well observed in soft X-rays
\citep{1992PASJ...44L.161S,1992PASJ...44L.173S}. 
Indeed all such jets, including
those in the quiet Sun, appear to have loop brightenings\index{jets!and loop brightenings} at their base. 
A jet appears as a collimated flow, implying the prior existence of large-scale or even
open magnetic fields\index{magnetic structures!jets and open fields}.
Type~III radio bursts\index{radio emission!type III burst!and jets}, known to be produced by electron
beams escaping from the corona into the solar wind, have a strong association
with the soft X-ray jets as well \citep{1994SoPh..155..203A,1995ApJ...447L.135K}.
Thus the microflare/jet events also have thermal and nonthermal attributes just as
major flares do, but the jet is relatively more prominent in the microflare domain.\index{jets!X-ray}

\subsection{Nanoflares and non-active-region phenomena}
\label{sec:hannah_nano}\index{quiet Sun}

We have seen above that the \textit{RHESSI} microflares occur only in active regions and
have a strong kinship with ``ordinary'' flares. How do these relate to X-ray bright
points \citep{1974ApJ...189L..93G}, network flares \citep{1997ApJ...488..499K},
plumes \citep{1978SoPh...58..323A}, or any of many more types of weak flare-like
brightenings seen both inside and outside the active regions?
Note that EUV
observations such as those of \textit{SOHO}/EIT \index{satellites!SOHO@\textit{SOHO}}
or \textit{TRACE}\index{satellites!TRACE@\textit{TRACE}} may be much more sensitive
than SXR (or HXR) observations of transient features at ordinary coronal
temperatures of order 1~MK \citep{1987ApJ...323..380P,1998ApJ...501L.213K}.
Many of these phenomena have been related to Parker's nanoflares and to the
problem of coronal heating\index{nanoflares!problem of energy scale}.
The nanoflare description of these EUV events is
justified, in the sense that an EUV brightening may have a much smaller event
energy than a microflare observed in SXRs, HXRs, or microwaves. 
However it is often difficult to place these different sorts of observations on a uniform energy scale\index{caveats!energy scale for occurrence distribution}.
We deal with this at length in Section~\ref{sec:hannah_distributions} below.

\begin{figure}\centering
\includegraphics[width=0.48\columnwidth]{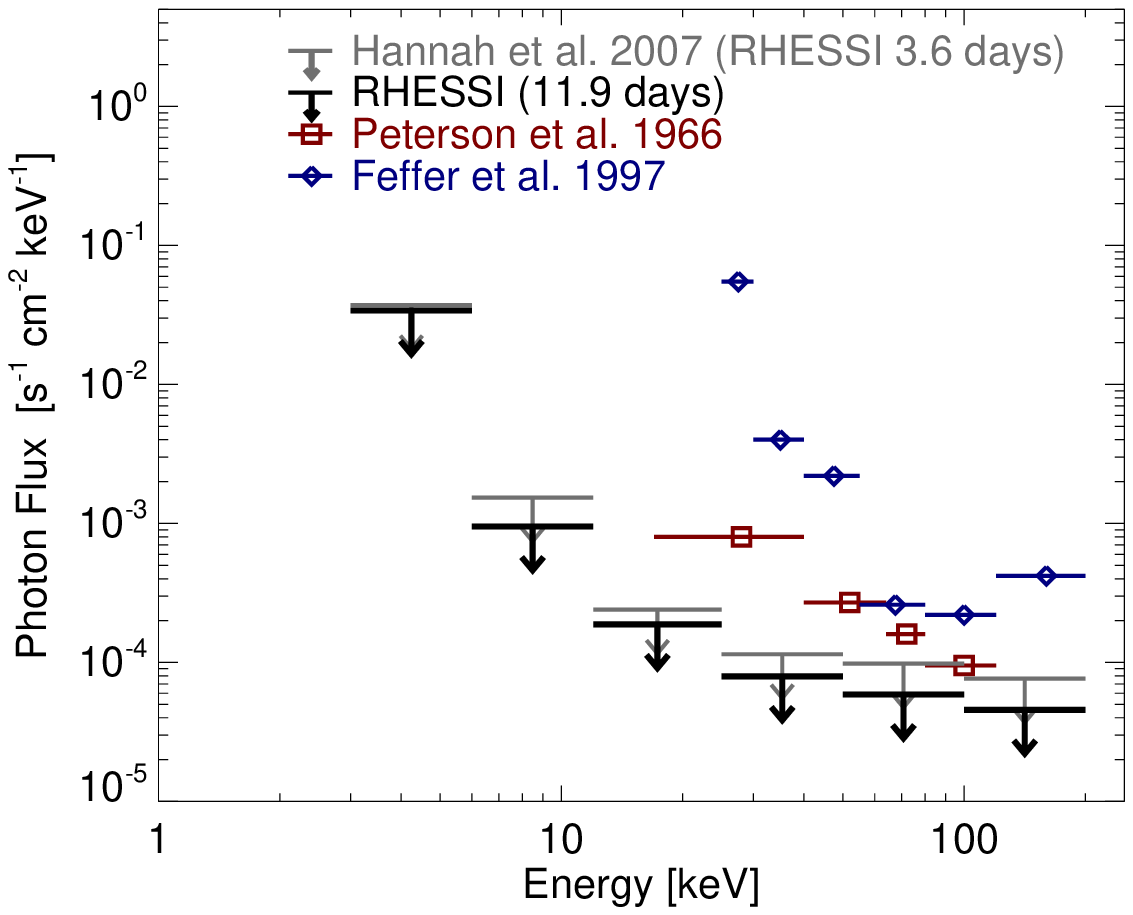}
\includegraphics[width=0.48\columnwidth]{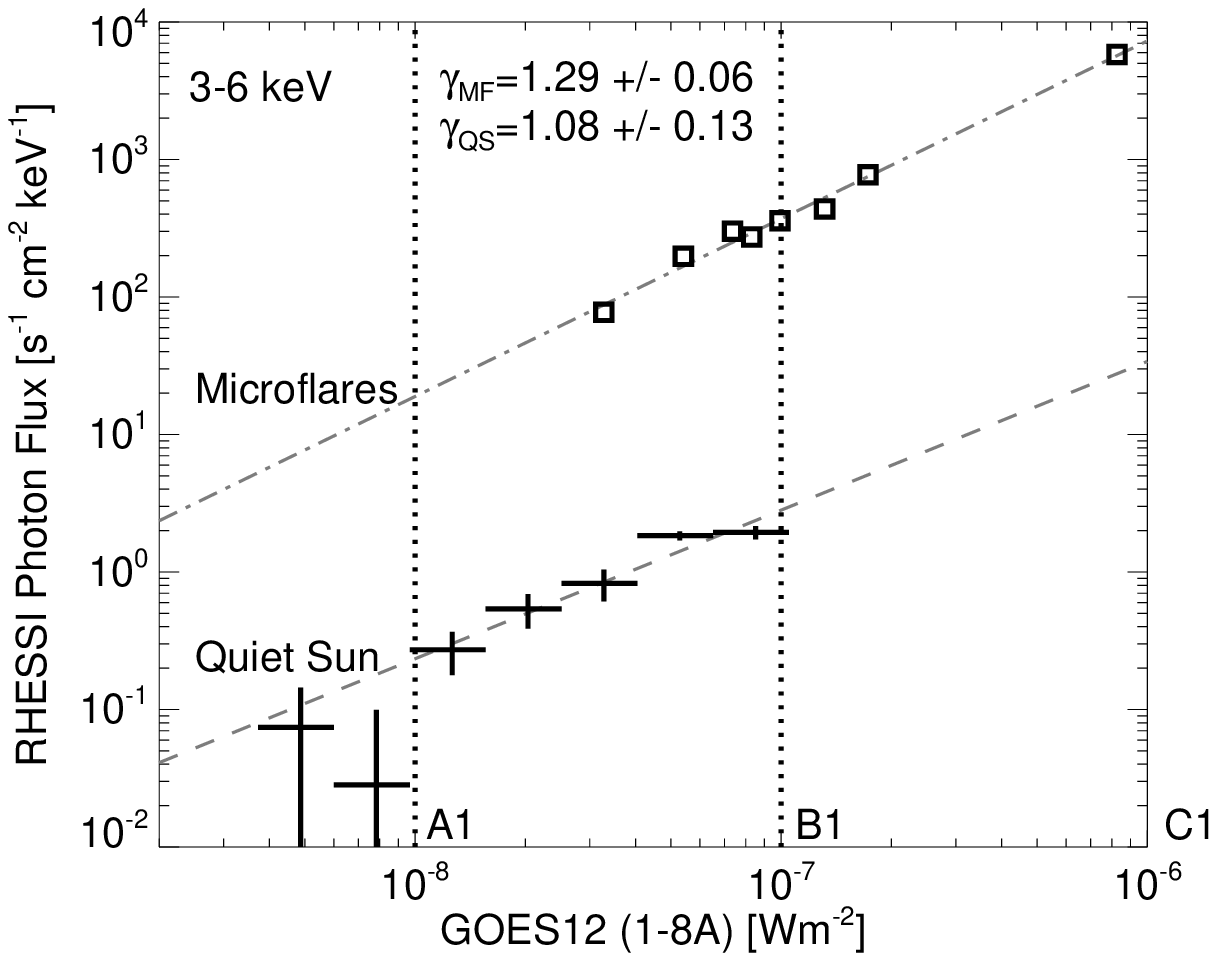}
\caption{
Observations and upper limits of the quiet-Sun hard X-ray flux found with
\textit{RHESSI}. (\emph{Left:}) the \textit{RHESSI} upper limits to the quiet Sun emission
at $3-200$~keV in comparison to previous observations
\citep{Hannah2010ApJ}, including the first \textit{RHESSI} quiet Sun limits
\citep{2007ApJ...659L..77H}. (\emph{Right:}) comparison of the $3-6$~keV flux
observed with \textit{RHESSI} to the $1-8$~\AA~ \textit{GOES} emission for the quiet Sun and
offpointed microflares. The $\gamma_\mathrm{MF}$ and
$\gamma_\mathrm{QS}$ indicate the index of power-law fits to the data, the
dashed and dashed-dotted lines
\citep{2007ApJ...659L..77H}. Reproduced by permission of the AAS. \label{fig:hannah_qs}}
\end{figure}
\index{hard X-rays!quiet Sun!illustration} 
\index{hard X-rays!quiet Sun}

Outside of active regions, the most prevalent X-ray features of SXR images are the
X-ray bright points (XBPs),\index{X-ray bright points (XBP)}\index{X-ray bright points (XBP)} originally discovered with rocket-borne X-ray imagers
in the late 1960s and studied statistically with data from the X-ray telescope on
\textit{Skylab}\index{satellites!Skylab@\textit{Skylab}} \citep{1974ApJ...189L..93G}.\index{Skylab@\textit{Skylab}}

Subsequent observations had sufficient
resolution to detect hot (about 2~MK) loop-like (about 10$''$  in length) properties
and footpoints in the XBPs\index{X-ray bright points (XBP)!footpoints}\index{X-ray bright points (XBP)!loops}
\citep{1994ApJ...430..913M,1992PASJ...44L.161S,2007PASJ...59S.735K}, pointing to a
possible relationship with flare physics. 
Typically hundreds of XBPs are visible,
uniformly spread across the solar disk \citep{1974ApJ...189L..93G}\index{X-ray bright points (XBP)!spatial distribution}, their number
changing little once observational bias, due to the dominant active region
emission, is removed \citep{2003ApJ...589.1062H}\index{magnetic field!XBP distribution}.
This spatially-uniform and
time-invariant distribution across the solar surface is expected as the XBPs are
associated with the magnetic fields related to surface convective flows\index{X-ray bright points (XBP)!and surface convective flows} (the granulation and supergranulation of the quiet photosphere) rather than with the
active-region fields responsible for sunspot fields.
\index{X-ray bright points (XBP)!and jets}
X-ray bright points are associated with connectivity changes in the network
magnetic field\index{sunspots!and network field}.
The underlying fundamental physical processes could be the same
as for active-region microflares. Often, XBPs may persist for hours to days, unlike
the transient active region flares, and thus their behavior may resemble that of
quiescent active-region loops \citep[e.g.,][]{1996ApJ...459..342Y}. In the absence
of solar active regions, the XBPs and the corona of the quiet Sun provide a basal
X-ray emission level\index{X-ray bright points (XBP)!and basal coronal emission}. 
Jets are sometimes observed from XBPs, particularly in polar
regions \citep{1992PASJ...44L.173S,2007PASJ...59S.745S,2007Sci...318.1580C,
2007PASJ...59S.771S} where there is easy access to open field lines.

There have been no HXR observations of XBPs.
\index{X-ray bright points (XBP)!hard X-rays}
Such emission therefore must be
below the sensitivity limits of current spacecraft, which have been optimized to
observe brighter active-region flares. Radio emission associated with XBPs has
been observed, but in many instances it is consistent with thermal instead of
nonthermal emission \citep{1992PASJ...44L.167N}.
\index{network flares}
Network flares, associated with
the magnetic network boundaries, are about an order of magnitude fainter than
XBPs and are more transient, lasting for only about 10 minutes
\citep{1997ApJ...488..499K}\index{gyrosynchrotron emission!rarity in network flares}.
Radio emission has also been observed in these events, but again in the majority of the cases it is consistent with thermal emission,  and in only a few cases could it be associated with nonthermal
gyrosynchrotron emission from accelerated electrons.

Determining whether there is nonthermal emission from a population of
accelerated electrons outside of active regions and in the quiet Sun would provide
important insights into the nature of possible small-scale steady-state
energization processes in the solar corona. 
\index{hard X-rays!quiet Sun}
\index{RHESSI@\textit{RHESSI}!HXR from quiet Sun}
Although \textit{RHESSI} has unprecedented
sensitivity over $3-25$~keV, which has greatly aided the study of small active-region
flares (see Section \ref{sec:hannah_micro}), emission outside active regions still
remains elusive.

Investigating the non-flaring properties of active regions
\citep{2009ApJ...697...94M} or the quiet Sun with \textit{RHESSI} is a non-trivial task.\index{active regions!non-flaring X-ray emission}
\textit{RHESSI}'s imaging method is designed for bright compact flares and ill-suited for
the spatially-diffuse emission expected from coronal heating. 
Instead, upper limits to the quiet Sun HXR flux\index{RHESSI@\textit{RHESSI}!HXR from quiet Sun}  can be obtain by offpointing\index{RHESSI@\textit{RHESSI}!offpointing} \textit{RHESSI} from the
Sun, so that the quiet solar signal is ``chopped'' as \textit{RHESSI} rotates
\citep{2007RScI...78b4501H}. 
This process has obtained limits that are smaller
and cover a wider energy range than previously found
\citep{2007ApJ...659L..77H,Hannah2010ApJ}, as shown in Figure~\ref{fig:hannah_qs}. A wide range of possible thermal and nonthermal emission
is still consistent with these limits with only a hint as to the relationship
between active region and non-active region flares.
\index{active regions!non-flaring X-ray emission}
However, the nanoflare model for coronal heating,\index{nanoflares!and coronal heating} operating in a manner similar to active-region flares, does seem unlikely \citep{Hannah2010ApJ}. 
Future instruments with higher sensitivity and dynamic range could possibly find such HXR emission, which would be expected from a nanoflare heating model even in the absence
of flare-like brightenings \citep[e.g.,][]{2004ApJ...605..911C}.
\index{hard X-rays!quiet Sun!and nanoflares}
\index{nanoflares}

\begin{center} \begin{table}
 \caption{\label{tab:hannah_flux}Table of the power-law indices $\alpha$ found
from statistical surveys of the fluxes of X-ray flares. The top set are predominantly
SXR surveys, the bottom HXR surveys.}

\begin{tabular}{cccccc} \hline\noalign{\smallskip} Index $\alpha$
& Energy Range & Duration & Number & Quantity &  Instrument \& Reference \\
\hline\noalign{\smallskip}
1.86& 7.7-12.5 keV & 1967(2 weeks) & 177 & Peak &
\textit{OSO-3} \citep{1969ApJ...157..389H} \\
1.75& 1-6.2 keV$^1$ & 1966-1968 & 4028 &Peak
&\textit{Explorer-33/35}
\citep{drake1971} \\
1.44& 1-6.2 keV$^1$ & 1966-1968 & 4028 &Fluence
&\textit{Explorer-33/35}
\citep{drake1971} \\
1.79& C{\sc A} XIX & 1980-1989 &  & Peak
&BCS \citep{lee1995} \\
1.86& 1.5-12.4 keV$^2$ & 1980-1989 &  & Peak
& \textit{GOES} \citep{lee1995} \\
1.88&1.5-12.4 keV$^2$ & 1993-1995 &1054  & Peak
&\textit{GOES} \citep{feldman1997} \\
2.11 & 1.5-12.4 keV$^2$ & 1976-2000 &49409  & Peak
&\textit{GOES} \citep{veronig2002} \\
2.03& 1.5-12.4 keV$^2$ & 1976-2000 &49409  & Fluence
&\textit{GOES} \citep{veronig2002} \\
\hline\noalign{\smallskip}
\hline\noalign{\smallskip}
1.8& 20 keV & 1971-1972 & 123  & Peak$^\mathrm{ph}$
&\textit{OSO-7} \citep{1974SoPh...39..155D} \\
2.0& 20 keV & 27-06-1980 &25  & Peak$^\mathrm{ph}$
&Balloon \citep{1984ApJ...283..421L} \\
1.8& $>$30 keV & 1980-1985 &$>$7000  & Peak
&HXRBS \citep{1985SoPh..100..465D} \\
1.66-1.75& $>$25 keV & 1980-1989 & $>$7000 & Peak
&BATSE \citep{1992como.work..457S} \\
1.61& $>$25 keV & 1991 & 1262 & Peak
&HXRBS \citep{1992como.work..457S} \\
1.75& $>$26 keV & 1978-1986 &4356  & Peak
&\textit{ICE} \citep{lee1993} \\
1.70-1.86& $>$30 keV & 1980-1984 &3578  & Peak
&HXRBS \citep{bai1993} \\
1.73& $>$25 keV & 1980-1989 &7045  & Peak
&HXRBS \citep{crosby1993} \\
1.59& $>$25 keV & 1980-1989 &2878  & Peak$^\mathrm{ph}$
&HXRBS \citep{crosby1993} \\
1.60-1.74& $>$25 keV & 1991-1994 &  & Peak
&\textit{CGRO} \citep{1994PhDT........51B} \\
1.86-2.00& $>$26 keV & 1978-1986 &3468  & Peak$^\mathrm{ph}$
&\textit{ICE} \citep{1995ApJ...455..733B} \\
1.74& $>$60 keV & 1980-1989 & 12327 & Peak
& HXRBS \citep{1997ApJ...475..338K} \\
1.58& $>$10 keV & 1989-1992 &1551  & Peak
&WATCH \citep{crosby1998} \\
1.56& $>$25 keV & 1991-1994 & 5430 & Peak
&BATSE \citep{1998ApJ...497..972A} \\
1.46& $>$50 keV & 1991-1994 & 5430 & Peak
&BATSE \citep{1998ApJ...497..972A} \\
1.50& 3-6 keV & 2002-2007 &24097  & Peak
&\textit{RHESSI} \citep{christe2008} \\
1.51& 6-12 keV & 2002-2007 &24097  & Peak
&\textit{RHESSI} \citep{christe2008} \\
1.58& 12-25 keV & 2002-2007 &24097  & Peak
&\textit{RHESSI} \citep{christe2008} \\
1.71& 4-8 keV & 2002-2007 &18656  & Peak$^\mathrm{ph}$
&\textit{RHESSI} \citep{hannah2008} \\

\hline\noalign{\smallskip} \end{tabular}\newline $^1$$2-12$~\AA, $^2$$1-8$~\AA.\\
The default units for the hard X-ray results are count rate (observed flux) except
those indicated by~$^\mathrm{ph}$ which are nominally instrument-independent
photon fluxes.
\end{table} 
\index{flare frequency distributions!observations of power laws!table}
\end{center}

\section{Flare distributions \& parameter scaling}\label{sec:hannah_obsdist}
\index{flare frequency distributions}

The occurrence distribution function provides one good way to characterize a large
number of observations. 
Many flare distribution functions have been published,
and we can separate them into two classes: distributions of raw observable
parameters (e.g., peak flux), and distributions of derived parameters (e.g., energy
content).
\index{flare frequency distributions!observed parameters!compared with derived}
A derived parameter such as the total event energy might seem a more
physically correct object, but the data manipulations required to get it will
necessarily involve uncertain calibrations and ``plausible'' assumptions about the
unobserved parts of the event. To make this latter point more concrete, consider
the two \textit{GOES} energy bands (0.5-4 \AA~ and 1-8 \AA). These define an isothermal
temperature and emission measure, but have little sensitivity at low coronal
temperatures (e.g., 1-2~MK). 
Thus in principle a large emission measure\index{caveats!emission-measure weighting} and
energy can be masked from view and go unaccounted for, systematically biasing
the energy estimate to be a lower limit. In the following we provide a critical
review of efforts to understand these different occurrence distribution patterns.

\subsection{X-ray flux distributions}\label{sec:hannah_distributions}
\index{flare frequency distributions!X-rays}

We summarize the reported X-ray flux and  fluence (the time-integrated flux)
distributions in Table~\ref{tab:hannah_flux}. Most of the power-law
indices~$\alpha$ fall below the critical value of~2.0, meaning that the energetic
events dominate, rather than the microflares (information on how to determine
$\alpha$ accurately is detailed in Appendix \ref{app:hannah_fitpowlaw}). An
exception to this is the huge \textit{GOES} event sample of \cite{veronig2002}, which we
showed in Figure~\ref{fig:hannah_goesdist}, but in this case the authors
deliberately did not attempt to correct for background counting rates\index{flare frequency distributions!background corrections}. 
Indeed, this is a fundamentally ambiguous procedure, as noted by \cite{1990ApJ...356..733B},
and a typical source of systematic error. The SXR observations suffer more from
this than the HXR observations because of their longer time scales\index{flare frequency distributions!temporal confusion}. 
Determining the fluence also introduces bias: how does one determine the time interval for
the integration, and how does one allow for sensitivity variations (either between
instruments or as a function of time)?

\begin{figure}\centering
\includegraphics[height=40mm]{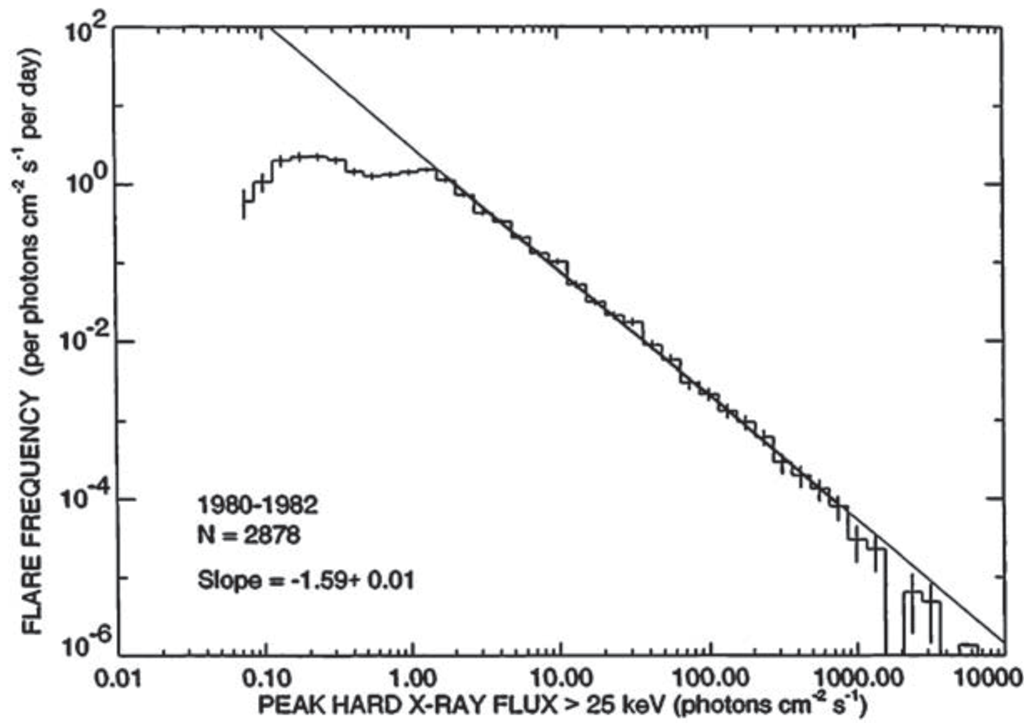}
\includegraphics[height=40mm]{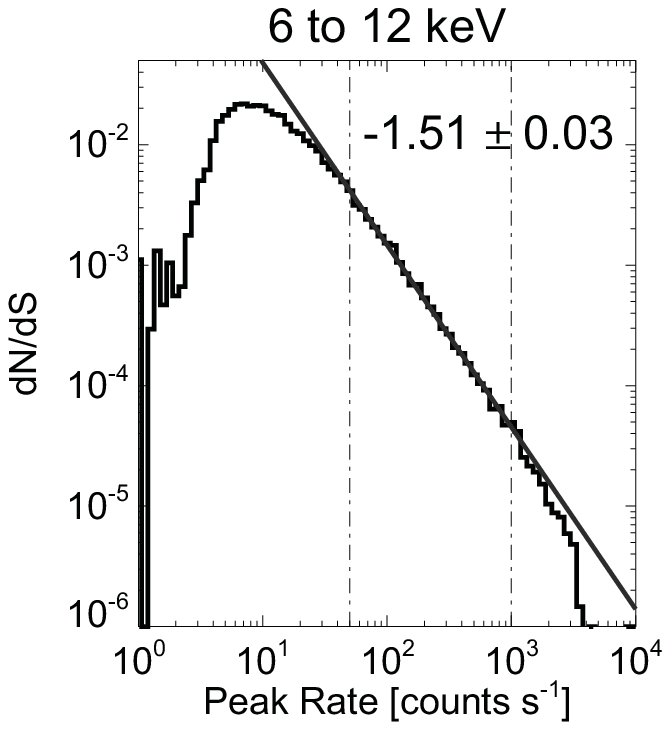}\\
\includegraphics[height=40mm]{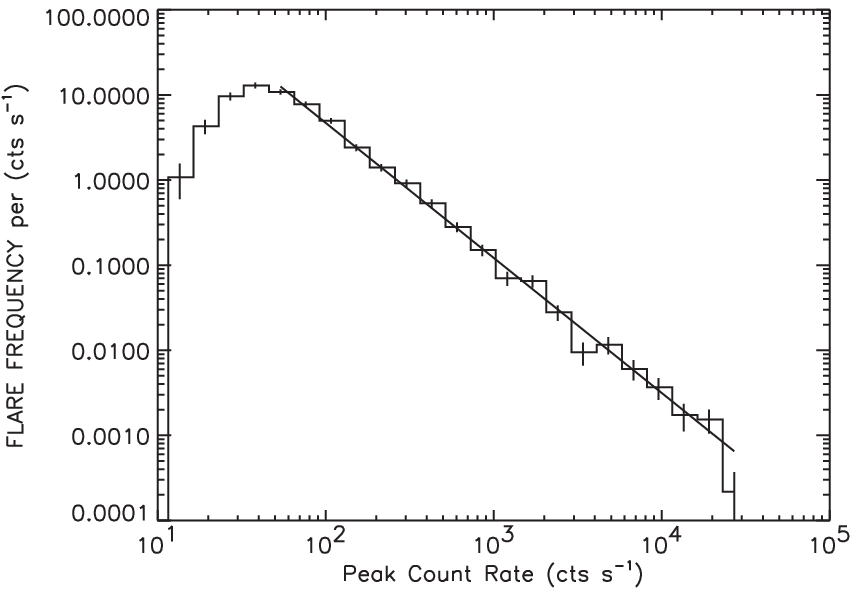}
\includegraphics[height=40mm]{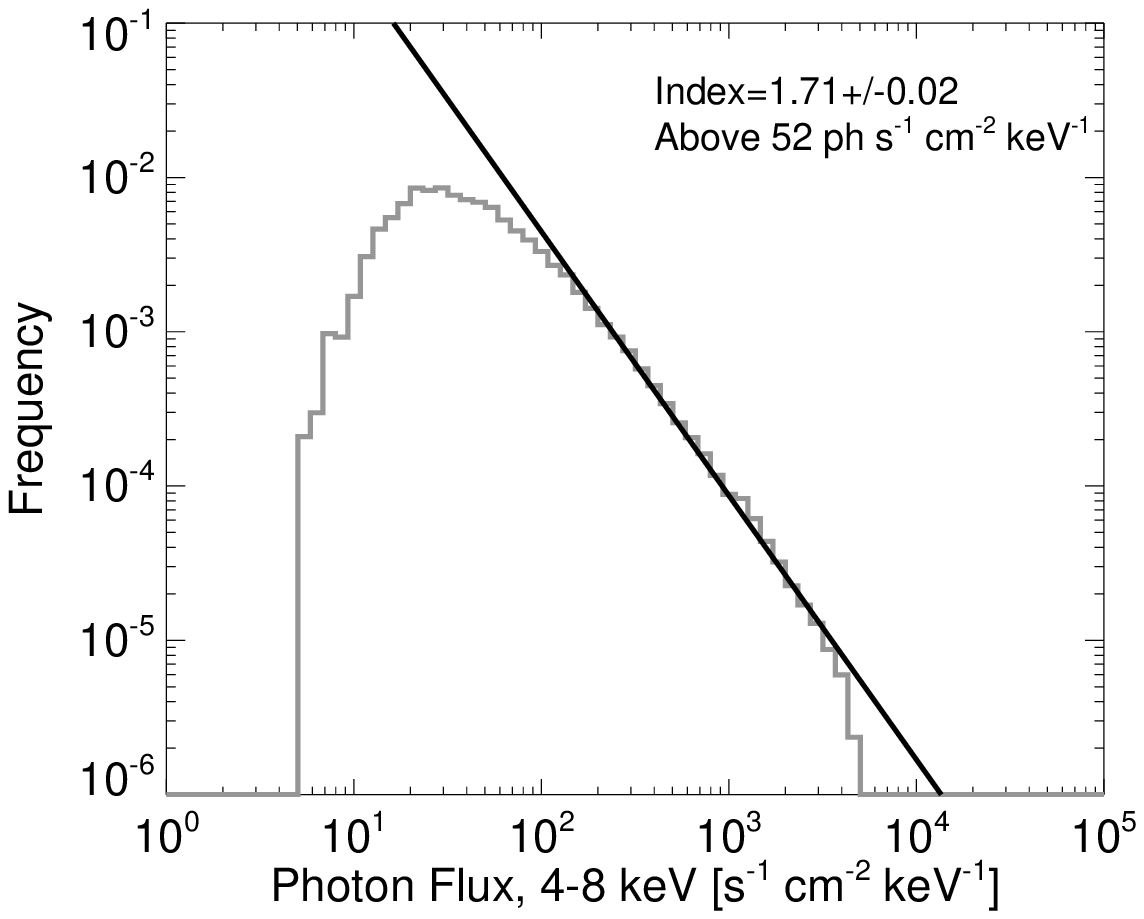}
\caption{\label{fig:hannah_fluxdist}Several of the flux distributions listed
in Table~\ref{tab:hannah_flux}
\citep{crosby1993,christe2008,crosby1998,hannah2008}
(left to right, top to bottom). Another example \citep{veronig2002} is shown in
 Figure \ref{fig:hannah_goesdist}. Reproduced by permission of the AAS.}
\end{figure}

Most of the entries in Table~\ref{tab:hannah_flux} (see also the representative
plots of several distributions in Figure~\ref{fig:hannah_fluxdist}) refer to
directly-observed peak fluxes. This is the simplest way to handle the data and one
that suffers the least from the introduction of systematic errors due to unknown
or imprecise corrections. Uncertainty due to background rates in the detector still
remains, as with the intrinsic biases of the sampling. This will have virtually no
effect on the largest events but dominate over the smallest ones. If the
instrument measures spectra in detail, it is also possible to take one step away
from instrumental bias by fitting a spectrum and then evaluating the spectral flux
density\index{flare frequency distributions!observed parameters!normalized to specific photon energy} at a well-measured photon energy, e.g., at 30~keV.

The representative distributions shown in Figure~\ref{fig:hannah_fluxdist} have
several common features: they match power laws well over a certain magnitude
range, and they roll off towards higher and lower magnitudes. The roll-off at the
low end is either due entirely to the sensitivity limit\index{flare frequency distributions!observed parameters!sensitivity limit} of the particular instrument,
or is heavily confused by selection effects due to this limit. 
In particular, faint events may be missed as their emission is obscured by brighter simultaneous
events. 
A fit to the distribution function that attempts to characterize this
roll-over \citep[e.g., a log-normal or a Weibull distribution;][]{parnell2002}
probably has little relevance to the physics of the events in the well-observed part
of the distribution. 
\index{distributions!log-normal} \index{distributions!Weibull}
At larger magnitudes the deficit also might have systematic
errors (the saturation of a given detector would be an obvious one), but at some
point a power-law fit for $\alpha  < $2 will diverge unphysically. This implies the
existence of an upper limit of some sort \citep[e.g.,][]{1980asfr.symp...69L}.
\citet{2007ApJ...666.1268A} analyzed 10 M- and X-class flares, finding that each was
composed of several bursts that saturated above 20 keV, suggesting a saturation of the HXR emission for electron-beams in large flares.
\index{hard X-rays!flux saturation}

The sensitivity and bias effects (whether instrumental or observational) result in a
clear change in the distributions at the extremes, but they can also have a more subtle
effect on the power law in the mid-regions.
\index{flare frequency distributions!biases} 
Although the power-law nature of the
distribution is not dramatically changed, the value of $\alpha$ can be different
\citep{2002ApJ...566L..59A,asch_parnell2002}. Some attempts have been made to
correct for these biases in the context of the derived energy distributions and will
be discussed further in Section \ref{sec:hannah_engdist}.

The distribution of event energies is obtained by integrating over the spectrum to
estimate an energy flux, and then over time\index{microflares!event energies!fluence determination}\index{flares!X-ray fluences}. Such conversions are not
model-independent. Therefore, the direct flux measurement may be a better
guide to the general conclusion from all such distributions: \emph{flare
occurrence is scale-invariant}\index{flares!scale invariance}. 
That is, the length scale does not change when
multiplied by a common factor,  a property of power-law distributions. Solar flares
thus have behavior resembling that of earthquakes\index{earthquakes} as described by the
\index{Gutenberg-Richter law} Gutenberg-Richter law \citep{gutenberg} but how general is it for flares? First, we
note that stellar flares\index{flares!stellar!occurrence distributions}\index{flare frequency distributions!stellar flares}, 
on a variety of stellar types, tend to follow similar
distributions \citep[e.g.][]{1989SoPh..121..375S}. 
From region to region, there can be slight variations in the distribution, in particular there being a varying upper
energy cutoff \citep{1997ApJ...475..338K}. 
\index{flare frequency distributions!active-region dependence}
This is lower for smaller active regions
as there is less free energy, but even in large regions there is a finite amount of
energy available (making ``super flares''\index{flares!super@``super flares''} not just infrequent but impossible). An
active region from a period of low solar activity has been found to have a
frequency distribution which clearly rolls over, this deviation from a power-law
being attributed to the low finite energy available \citep{2010ApJ...710.1324W}.

\begin{figure}\centering
\includegraphics[height=50mm]{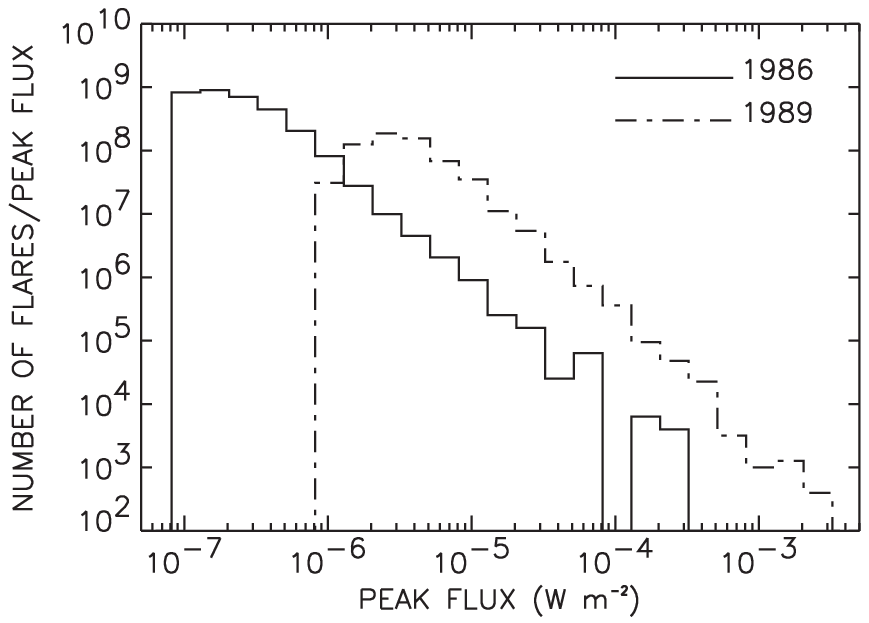}
\includegraphics[height=50mm]{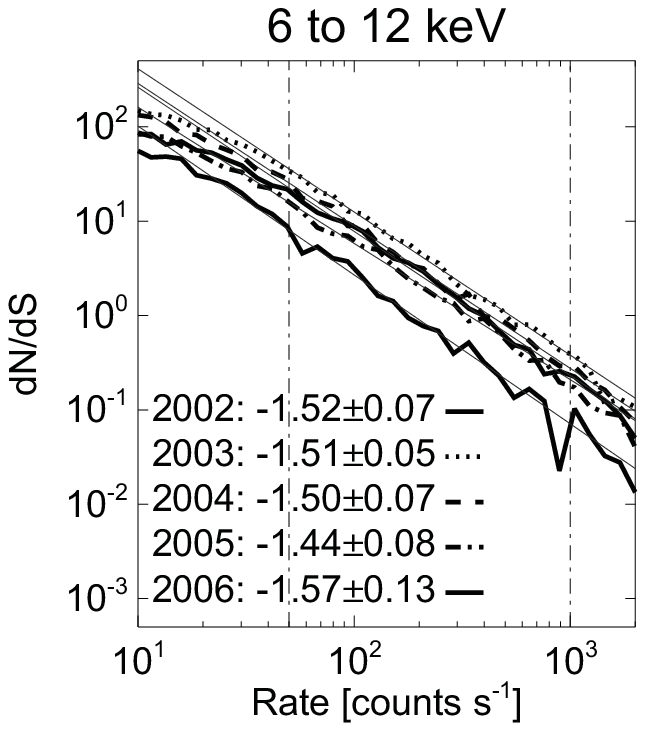}\\
\caption{
\label{fig:hannah_distperyear} Comparisons of occurrence distributions with
phase in the solar cycle. (\textit{Left}) \citet{veronig2002} using \textit{GOES} 1-8
\AA~data from solar minimum (1986) to maximum (1989); (\textit{right})
\citet{christe2008} using \textit{RHESSI} microflare data from leaving solar maximum
(2002) through to approaching solar minimum (2006).
Reproduced by permission of the AAS.} \end{figure}

There can also be variations on intermediate time scales
\citep{bai1993,1995ApJ...455..733B}. However, in general, these are not strong
deviations from the general pattern. From \textit{RHESSI} we find
(Figure~\ref{fig:hannah_distperyear}, right panel) little evidence for change in the
slope of the microflare distribution as a function of phase of the solar cycle, with
excellent statistical significance. This is consistent with the results of
\cite{veronig2002} for \textit{GOES} events (Figure~\ref{fig:hannah_distperyear}, left
panel). 
\index{scale invariance}
The scale-invariant property of the flare occurrence distributions is
thought to provide the evidence that the coronal magnetic field is in a
self-organized critical state \citep{1991ApJ...380L..89L} and this is detailed further
in Section \ref{sec:hannah_whyapowlaw}.
\index{self-organized critical state}

\subsection{Time distributions}
\index{flare frequency distributions!durations}

\begin{figure}\centering
\includegraphics[height=45mm]{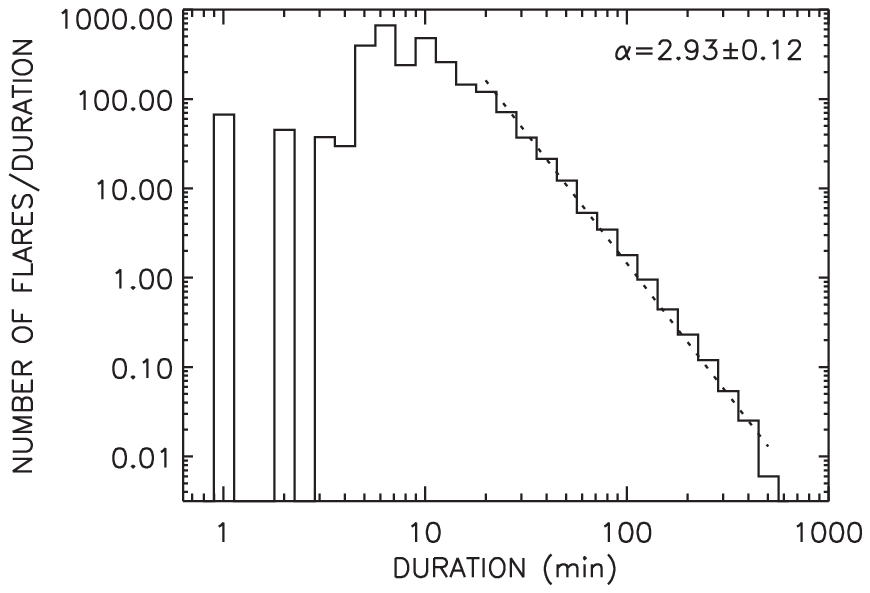}
\includegraphics[height=45mm]{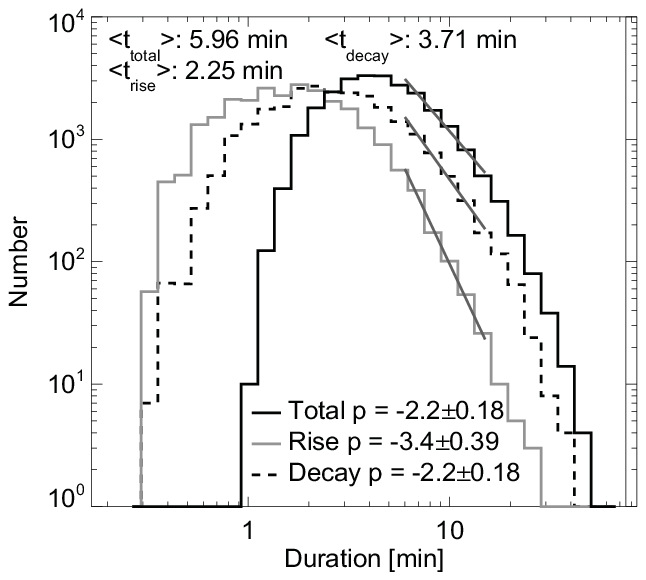}
\caption{
\label{fig:hannah_distduration}
\textit{Left:} the distribution of the durations of flares \textit{GOES} $>$B class 
\citep{veronig2002}; \textit{right:} distribution of durations of
microflares \citep{christe2008}, typically below \textit{GOES} C~class.
Reproduced by permission of the AAS.}
\end{figure}

The duration of an event must be known for an estimation of the fluence and
therefore also for an estimation of the event energy.
Figure~\ref{fig:hannah_distduration} shows two examples of duration
distributions: from \textit{GOES} \citep{veronig2002} and from \textit{RHESSI} \citep{christe2008}.
The \textit{GOES} data have a clear power-law falloff to long durations \cite[see
also][]{1971SoPh...16..152D}, whereas the \textit{RHESSI} distributions (shown separately
for rise, fall, and total times) have more symmetrical distributions. This is because
the \textit{RHESSI} data are considerably more affected by selection bias. Trying to
determine the duration of the flare crucially depends on being able to distinguish
the flare from the background. The \textit{RHESSI} distribution shown is for microflares
(smaller than \textit{GOES} C-class) for which it is difficult to separate their start and end
times from the background. The \textit{GOES} distribution is only for B-class events and
larger ($>$10$^{-7}$W~m$^{-2}$) and the background rate in \textit{GOES} varies less than in
\textit{RHESSI}.

\cite{lee1993} explicitly deal with the simultaneous flux and duration distributions
\index{flare frequency distributions!multivariate analysis} 
of an HXR data set, that of (\textit{ICE}\index{satellites!ISEE-3@\textit{ISEE-3}}\index{satellites!ICE@\textit{ICE}},
formerly \textit{ISEE-3}) above 26~keV \citep{1978ITGE...16..157A}. This analysis of joint
variables requires an explicit consideration of the parameter-space domain as a
means of understanding sampling bias. \cite{lee1993} concluded that a correlation
between the duration and peak flux was not present; this would be a requirement
if microflares or nanoflares were to outweigh the energy present in the ordinary
flares. 
Thus, the power laws observed in the distributions of peak fluxes most
likely provide a guide to the ``true'' distribution of total event energies.

The other important temporal signature of flares is the time between one flare
and the next one, the \emph{waiting-time}.
\index{waiting-time distribution}
This is generally taken to be the time
between the peak emission of one flare and the next. The waiting-time
distributions provide information about the probability of a flare occurring. 

These distributions can be predicted by self-organized criticality
\index{self-organized critical state} 
\index{flare models!avalanche}
\index{flare models!self-organized criticality (SOC)}  (avalanche) models of the coronal
magnetic fields, discussed further in Section~\ref{sec:hannah_whyapowlaw}.
These avalanche models have an exponential waiting-time distribution which
corresponds to a constant Poisson flare occurrence rate
\citep{1998ApJ...509..448W}. Such a flare occurrence model produces the desired
power-law distribution of energy and duration. Individual active regions
demonstrate waiting-time distributions that are either exponentials or the sum of
exponentials \citep{2001SoPh..203...87W,2001JGR...10629951M}. The latter results
can be explained by the Poisson occurrence rate varying as the active region
crosses the solar disk.

\begin{figure}\centering
\includegraphics[height=40mm]{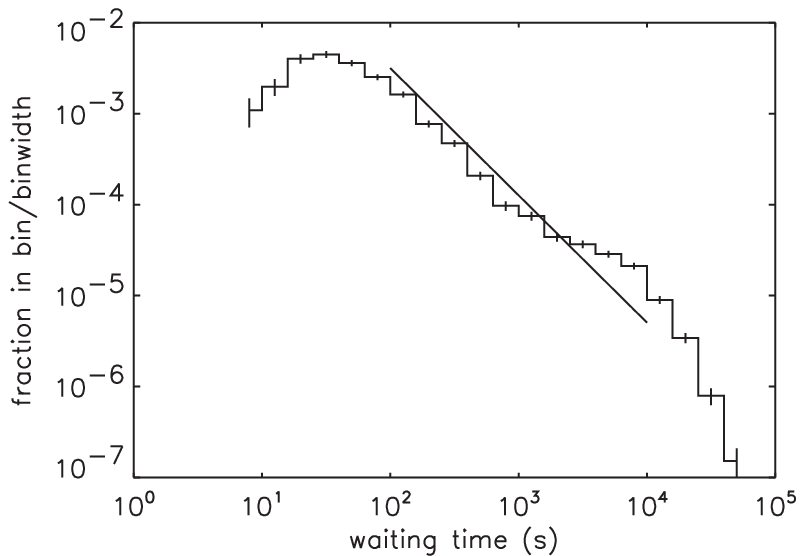}
\includegraphics[height=40mm]{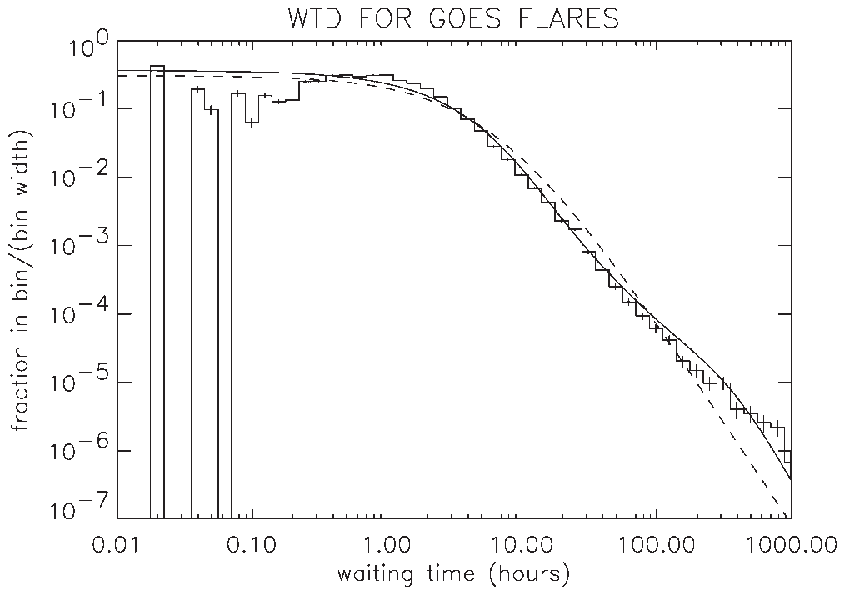}
\caption{\label{fig:hannah_wtd}
Waiting-time distributions for different sets of X-ray flares. \emph{Left:}
\textit{ICE/ISEE-3} study of 6919 flares $>$30~keV \citep{1998ApJ...509..448W}, the
overplotted solid line is a power-law fit. 
\emph{Right:} 32,563 \textit{GOES} flares
above C-class \citep{2000ApJ...536L.109W}, the overplotted lines indicate a
model for a time-dependent Poisson process (solid) and independent events
with an exponential distribution of rates (dashed). Reproduced by permission of the AAS.}
\end{figure}

The waiting-time distribution\index{waiting-time distribution} 
found for large samples of X-ray flares is shown in
Figure \ref{fig:hannah_wtd}. The sample of 6,919 flares observed $>$30~keV with
\textit{ICE/ISEE 3} is found to have a waiting-time distribution that is neither power-law
nor exponential \citep{1998ApJ...509..448W}, shown left in Figure~\ref{fig:hannah_wtd}. 
A larger study of 32,563 \textit{GOES} C-class and above flares
demonstrates a power-law tail in its waiting-time distribution over long
timescales \citep{2000ApJ...536L.109W}, shown right in 
Figure~\ref{fig:hannah_wtd}.  
The index of this power law varies with the solar cycle
\citep{2002SoPh..211..255W}, again consistent with a Poisson occurrence with a
time-varying rate.
\index{waiting-time distribution!time-varying Poisson occurrence} 
An alternative model explains this tail using an occurrence rate
with a L\'{e}vy distribution \citep{2001ApJ...555L.133L} but this also requires a
``memory'' in the underlying process. 
This suggests that not only can flare rates be
determined but features of the underlying physical processes can be understood, although this is still
under debate. 
The waiting-time distribution  of Type III radio bursts from an active region have been found to be consistent with a Poisson process \citep{2010ApJ...708L..95E}.
\index{radio emission!type III burst!waiting time distribution}
\index{waiting-time distribution!type III bursts}

An overabundance of short waiting-times compared to simulations has now been
found \citep{1998ApJ...509..448W}, suggesting that HXR bursts are not
independent events. 
This \emph{sympathetic flaring} behavior has frequently been
suggested previously \citep{1976SoPh...48..275F}, referring to temporally close
flares in different active regions. 
This may also relate to the misidentification of several peaks within a single flare as multiple events. 
It is difficult with non-imaging instruments to exclude closely related flares from the
same active regions, but statistically significant evidence for sympathetic flaring
has been found \citep{2002ApJ...574..434M}. \textit{RHESSI} would be able to provide such
information, as well as investigating waiting-times for smaller HXR flares, but the
highly discontinuous nature of the data (with gaps due to nighttime, South
Atlantic Anomaly passage, etc.) would make such analysis highly subject to selection effects.
\index{caveats!gaps in time series}

\subsection{Origin of power-law distribution}\label{sec:hannah_whyapowlaw}
\index{flare frequency distributions!why a power law?} 

An early analytic model for the statistics of solar-flare occurrence assumed that the
available energy grew exponentially with time and would be released as a flare
with a Poisson probability distribution
\citep{1978ApJ...222.1104R}. 
\index{flare frequency distributions!and Poisson distribution} 
This results in a power-law frequency distribution. 
This problem was later recast as a steady-state
transport equation, allowing the inclusion of an arbitrary energy resupply rate
\citep{1994SoPh..151..195L}. These models assume that all of the free energy is
released in a flare, as in a relaxation oscillator\index{relaxation oscillator}. 
Such behavior has never been found in the occurrence patterns of solar flares, although there are other astrophysical contexts in which it has \citep[e.g.,][]{1976ApJ...207L..95L}.
\citet{1998ApJ...494..858W} and \citet{2008ApJ...679.1621W} generalized these
models to take into account that the flares do not release all of the free energy by
finding a ``master equation''\index{master equation} to describe the system. 
This model again produces a power-law flare frequency distribution as well as a high-energy turnover.

A complementary view is that the scale-invariant\index{scale invariance} behavior of the flare
distributions implies that the system dynamics can be described as variations
around a self-organized critical state
\index{self-organized critical state}
\citep{1991ApJ...380L..89L,1993ApJ...412..841L}.
In the standard view, magnetic
energy builds up in complicated and stressed magnetic field structures in the
corona due to the motion and emergence of magnetic fields through the
photosphere. 
Eventually, the coronal structure loses equilibrium catastrophically\index{loss of equilibrium},
and its restructuring (usually thought to involve magnetic reconnection) suddenly
liberates some of the built-up magnetic energy. This newly released energy goes
into the various forms we observe in a solar flare. In the scenario of the
self-organized critical state, the instabilities are spontaneous, independent of the
history of how the energy accumulated, and directly trigger a cascade of energy
releases, ending in a temporarily stable state. 
Therefore, the same process could
easily produce events at the nanoflare \index{nanoflares!and cellular automata}level as well as the more energetic flares.
The avalanche-like behavior of such a system is frequently described using an
idealised ``cellular automaton'' model\index{flare models!cellular automaton}\index{cellular automaton}
\citep{1986JGR....9110412K,1987PhRvL..59..381B}.
Such a model does not depend
upon the actual physical mechanisms involved at the microscopic level, replacing
them with a schematic set of \textit{ad-hoc} rules for the system evolution. The links
between the physics of such a model and its statistical description remain unclear,
though taking a steady-state energy release in the ``master equation''\index{master equation} is similar to assuming an underlying avalanche process \citep{1998ApJ...494..858W}.
Nevertheless an extensive literature applying cellular automata to solar flares has
arisen \citep[e.g.,][]{2001SoPh..203..321C}.

\subsection{Nonthermal and thermal spectral parameters}\label{sec:hannah_tnt}
\index{microflares!X-ray spectral parameters}

The \textit{RHESSI} X-ray spectra bridge the thermal and nonthermal spectral domains, at
the same time providing better spectral resolution than any earlier HXR imager.
The nonthermal emission of the impulsive phase is often characterized by brief
spikes of emission with soft-hard-soft\index{hard X-rays!spectra!soft-hard-soft} (steep to flat to steep) spectral evolution, and its simplest characterization is a power law in energy.\index{soft-hard-soft}
The SXR thermal emission roughly follows the time integral of the HXR impulsive-phase emission,
following the Neupert effect\index{soft X-rays!Neupert effect!microflares} \citep{1968ApJ...153L..59N}.
This empirical result\index{Neupert effect} is
thought to demonstrate that the total nonthermal energy deposited heats the
chromospheric plasma, driving it up into flaring loops, producing bright thermal
emission.

\begin{figure}\centering
\includegraphics[height=35mm]{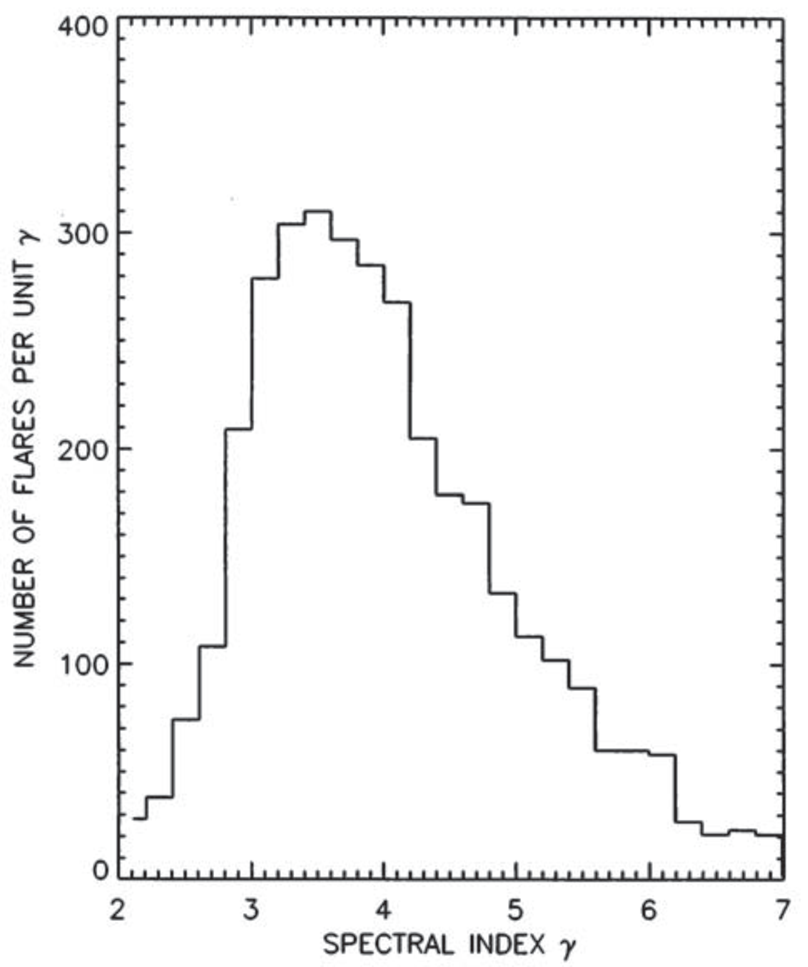}
\includegraphics[height=35mm]{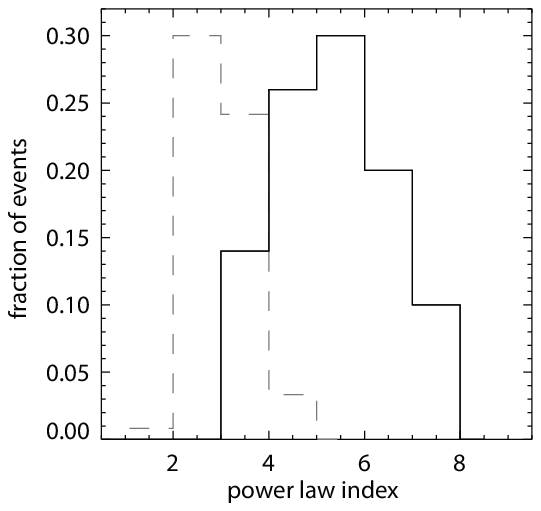}
\includegraphics[height=35mm]{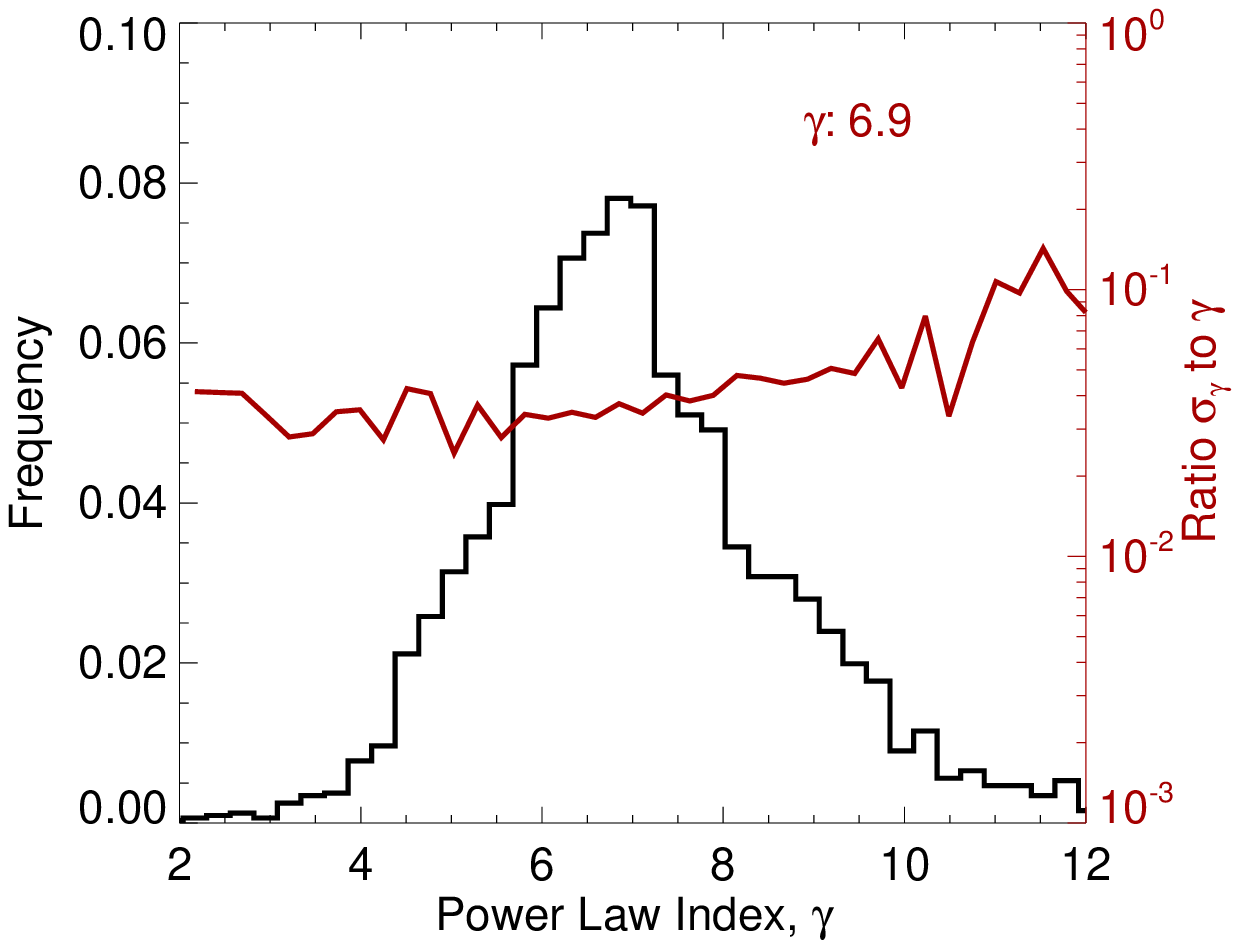}\\
\caption{
\label{fig:hannah_gamdist} Distributions of HXR power-law index $\gamma$ for
\emph{Left:} 6919 large flares observed above 30keV with \textit{ICE/ISEE-3}
\citep{1995ApJ...455..733B}; \emph{middle;} 55 large flares on the disk (dashed
grey) and occulted (solid black) with \textit{RHESSI} \citep{2008ApJ...673.1181K}.
\emph{Right:} 9161 \textit{RHESSI} microflares from the sample of  \cite{hannah2008},
the dark red line indicating the error ratio. Reproduced by permission of the AAS.}
\index{occulted sources!illustration}
\end{figure}

\begin{figure}\centering
\includegraphics[height=40mm]{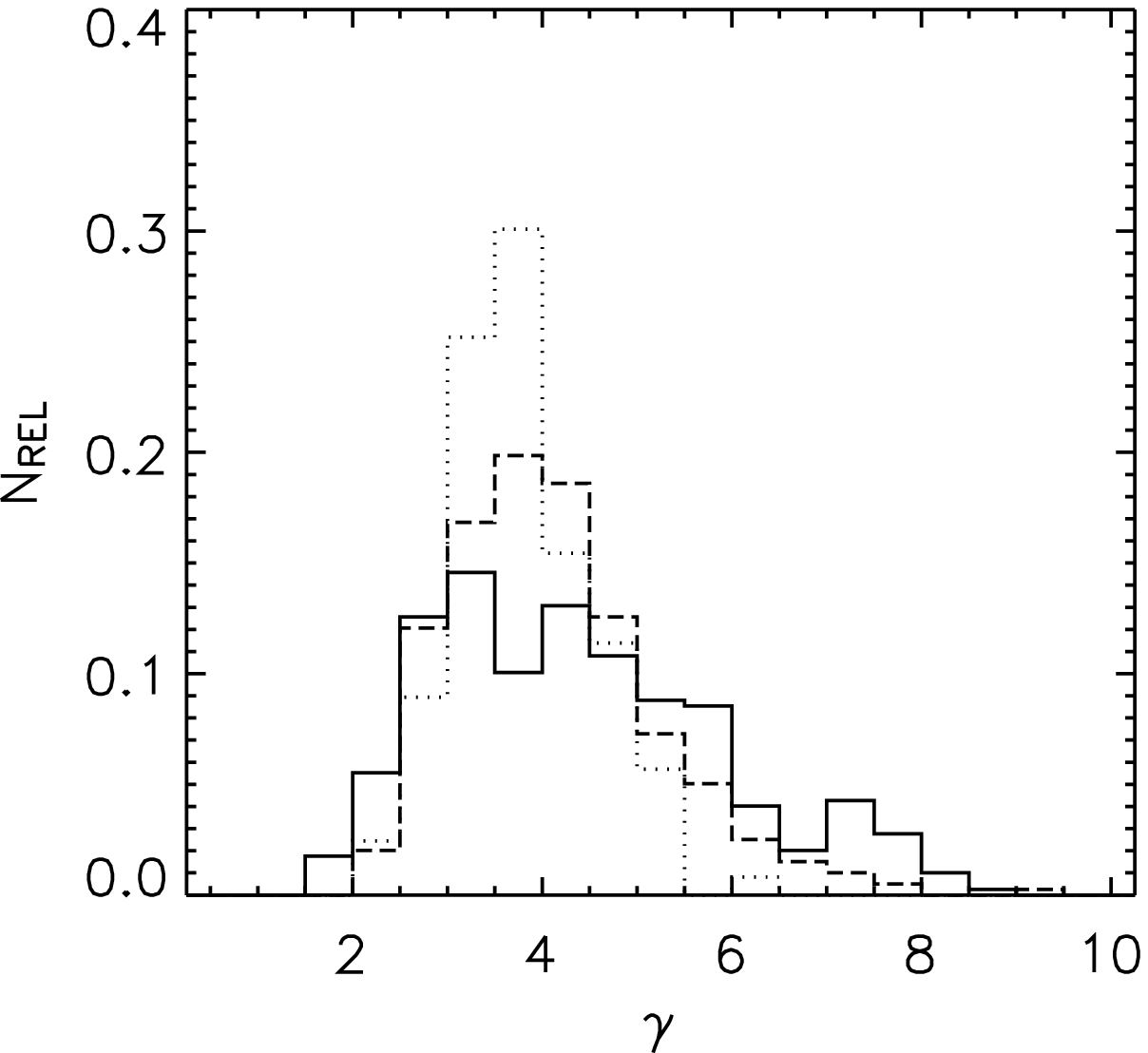}
\includegraphics[height=40mm]{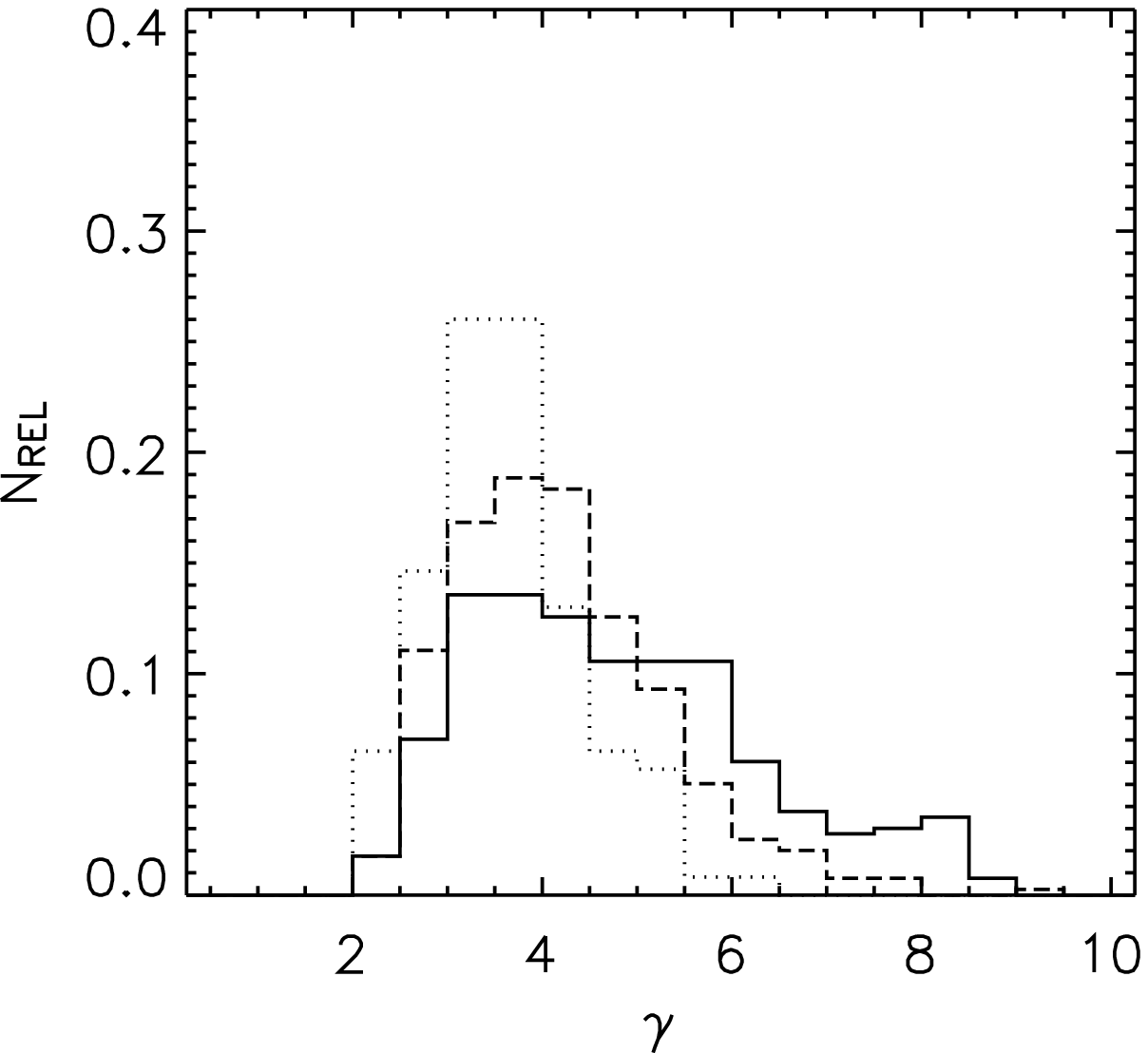}\\
\caption{
\label{fig:hannah_janagamdist}The effect of albedo correction\index{albedo} on the observed
HXR power-law index, with histograms of $\gamma$ (\emph{left}) found
from spectrum fitting  and (\emph{right}) then corrected for albedo
\citep{2007A&A...466..705K}. The different lines show the power-law fits made
over different energy ranges: 398 flares with $\gamma_0$ (solid) 15-20 keV and
$\gamma_1$
(dashed) 20-35 keV, and 123 flares with $\gamma_2$ (dotted) 35-50 keV. }
\end{figure}
\index{albedo!correction!illustration}

\subsubsection{Nonthermal}
\index{electrons!nonthermal}

As regards energetics, the HXR parameter of interest is the slope of the power-law
fit to the photon spectrum in the impulsive phase, namely $\gamma$ in
$I(\epsilon)\propto \epsilon^{-\gamma}$. This spectrum gives us the best
possible guide to the behavior of the energy in the impulsive-phase electrons\index{electrons!spectrum!selection effects},
which can be highly significant \citep{1971SoPh...18..489B,1976SoPh...50..153L}.
Figure~\ref{fig:hannah_gamdist} shows the distributions of the power-law
index~$\gamma$ from several data sets. 
Each shows clearly distinguishable selection effects. 
The left panel of Figure~\ref{fig:hannah_gamdist} displays the
spectral index of large flares above 30~keV as found from \textit{ICE/ISEE-3} (pre-\textit{RHESSI}).
\textit{RHESSI} observations of large flares indicate similar spectral indices, with occulted
flares (those with their footpoints hidden behind the limb so the HXRs are coronal in
origin) have systematically softer spectra \citep{2008ApJ...673.1181K} as seen in
the middle panel of Figure~\ref{fig:hannah_gamdist}. \textit{RHESSI} microflares
(Figure~\ref{fig:hannah_gamdist}, right panel) show many much steeper spectra
which are often difficult to distinguish from the thermal emission
\citep{hannah2008}.\index{occulted sources}
\cite{2007A&A...466..705K} were able to isolate one
important source of systematic error, namely the albedo\index{albedo!correction}\index{albedo!microflare spectra}\index{microflares!albedo correction} resulting from X-rays backscattered off the photosphere \citep{1973SoPh...29..143S,chapter7}. 
The albedo correction affects mainly low energies, $\sim$20~keV. 
The result of the correction is a steeper photon spectrum at those energies, as 
shown in detail in Figure~\ref{fig:hannah_janagamdist}. 
\textit{RHESSI} allows imaging spectroscopy of
individual HXR footpoints and the difference in spectral index between footpoints
in large flares (above \textit{GOES} M-class) with two well-resolved footpoints ranges
between 0 to 0.6 \citep{2008SoPh..250...53S}.

\begin{figure}\centering
\includegraphics[height=45mm]{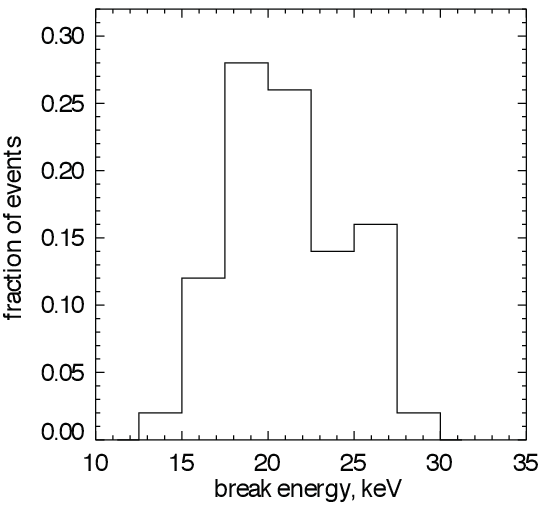}
\includegraphics[height=45mm]{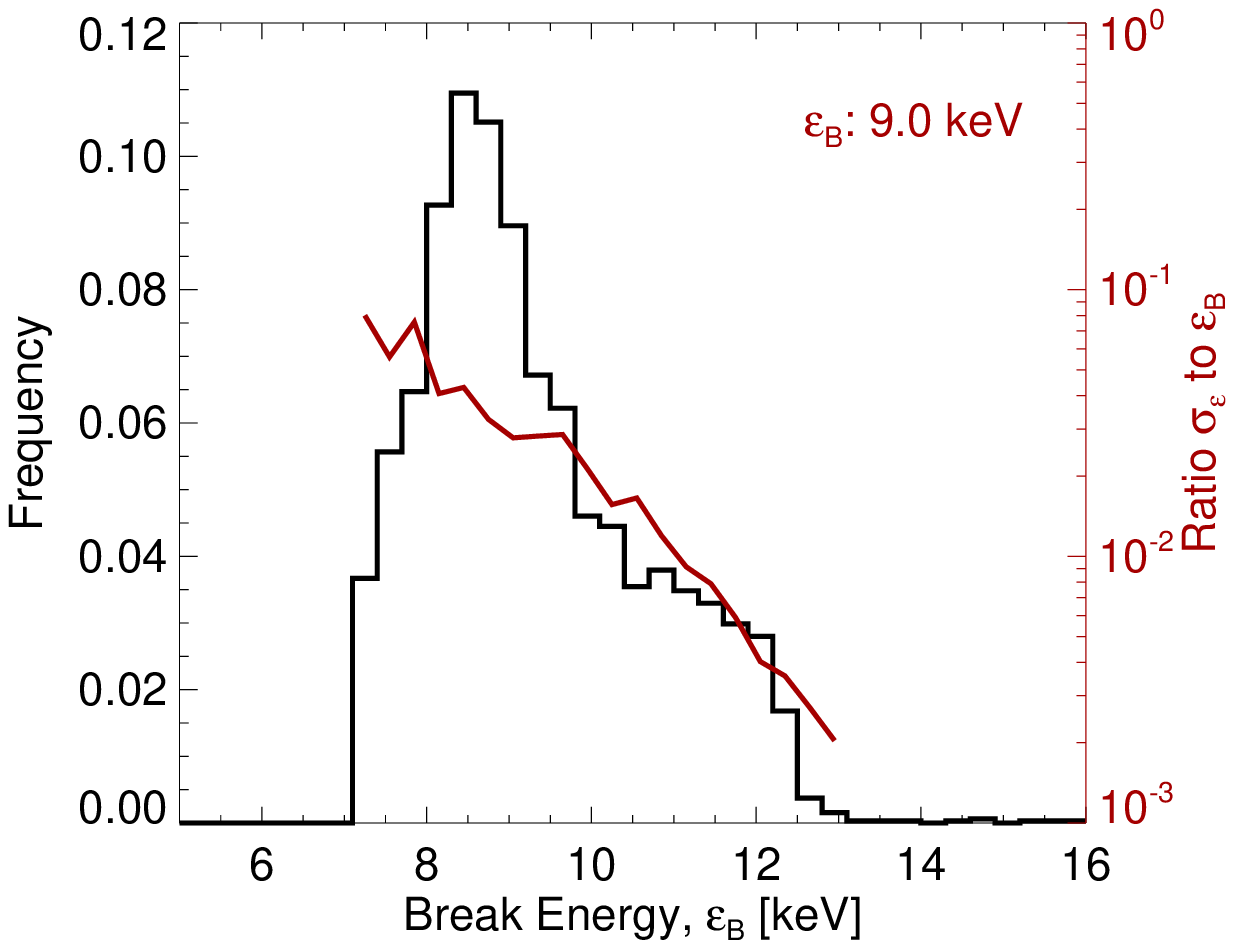}\\
\caption{
\label{fig:hannah_ebdist} Two examples of low-energy break energy
distributions from fitting \textit{RHESSI} photon flare spectra, (\emph{left}) for 55
large occulted flares \citep{2008ApJ...673.1181K} and (\emph{right}) 4236 microflares
\citep{hannah2008}, the dark red line indicating the error ratio.
Reproduced by permission of the AAS.}
\index{occulted sources!illustration}
\end{figure}

The essential problem in the interpretation of the hard X-ray spectrum lies in the
behavior of the bremsstrahlung cross-section \cite[e.g.,][]{1971SoPh...18..489B}.
\index{bremsstrahlung}
The photon spectrum is an integral over the electron spectrum at  all higher
energies. This smearing effect means that, in practice, it is difficult to determine
the parent electron spectrum at low energies, and yet these lower-energy
electrons contain most of the energy supplied to the flaring plasma via collisions.
Analytically, the electron spectrum can be assumed to exist above some ``cutoff'' energy
$E_C$, with means the accelerated electron population appears abruptly at
energies where the thermal spectrum is negligible. 
\index{low-energy cutoff} 
This results in a flattening of
the expected photon spectra, with a ``break'' energy $\epsilon_B$ occurring at photon
energies below the electron cutoff energy \citep{1988ApJ...324.1118K}.
\index{low-energy cutoff!in microflare hard X-ray spectra}
Although the spectral indices of the source electron distributions and observed photon
spectra are related in a simple way \citep{1971SoPh...18..489B,chapter7},
this is not the case for the photon break and electron cutoff energies\index{inverse problem!and microflare hard X-ray spectra}.
There are large and
ill-defined uncertainties in the inverse or forward deconvolutions of the HXR
spectrum for these parameters.

\begin{figure}
\centering\sidecaption
\includegraphics[height=55mm]{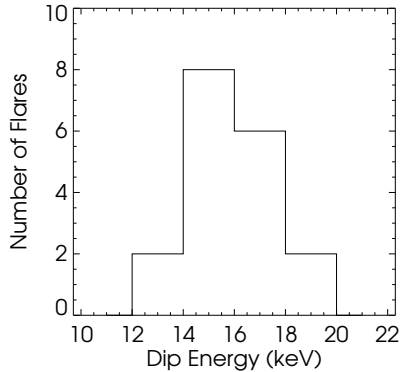}
\caption{\label{fig:hannah_dipdist} Histogram of the measured local minimum of
the \emph{dip} in the mean electron flux spectrum for a sample of 18 flares
\citep{2008SoPh..252..139K}. These dips can be removed by correcting for
albedo and do not represent a low-energy cutoff.}
\end{figure}

\cite{2007ApJ...670..862S} discuss the behavior of the break energy in a selection
of ``early impulsive'' events, which have less confusion between thermal and nonthermal  components.
\index{flare types!early impulsive}
\index{electrons!nonthermal!early impulsive events}
They found that 9~out of~33 early impulsive flares
demonstrated spectral flattening in the X-ray spectrum towards low energies, a
signature of a low-energy cutoff in the electron spectrum. 
In three of these events, albedo correction removed the flattening, but in the remaining six, the flattening was found to be consistent with forward-fitting a low-energy cutoff in the range 
$15-50$~keV, which also correlates with the HXR flux. This is at lower energies than
found from fitting the X-ray spectrum of flares without early impulsive emission.

Prior to \textit{RHESSI} this confusion was not a problem, as an instrumental cutoff of
around $20-30$~keV meant that the transition of the nonthermal to thermal
components of the spectra was rarely observed. Figure~\ref{fig:hannah_ebdist}
shows two examples of break-energy distributions\index{flare frequency distributions!break energy} derived from the \textit{RHESSI} data,
the left panel showing large occulted flares \citep{2008ApJ...673.1181K} and the right
panel showing microflares \citep{hannah2008}.\index{occulted sources}
For microflares, the observed
break tends down to very low energies where there are multiple emission lines in
the thermal spectra, as can be seen in Figure~\ref{fig:hannah_mfspec}. This makes
it very difficult to determine $\epsilon_B$ accurately. In addition, a model
requiring a sharp cutoff in the nonthermal electron distribution down at low
energies where the thermal spectrum is non-negligible does not seem
appropriate, as a smoother transition is expected. The validity of this model and
the effort to determine the energy in a microflare's nonthermal electrons  is
problematic and is discussed in detail in Section\ref{sec:hannah_engdist}.

Instead of trying to fit models to the photon spectrum based upon the expected
analytical derivation from the electron distribution, \citet{2008SoPh..252..139K}
inverted the photon spectrum directly to obtain the mean electron spectrum for
several flares.
They found that there was no sharp cutoff \index{electrons!spectrum!low-energy cutoff} in the electron distribution, but instead a dip between $12-20$~keV.
A histogram of these dip energies is shown in Figure~\ref{fig:hannah_dipdist}. 
Making albedo corrections\index{albedo!dip in spectrum}  to these spectra completely removed the dip, suggesting that there is a smooth transition between the thermal and nonthermal emissions. 
Their study also suggests that if low-energy cutoffs exist in the mean electron spectra of these
flares, they should be located at energies less than $\sim$12~keV. 
This implies that applying a cutoff model is only appropriate when considering an electron
population well away from the thermal distribution.
\index{caveats!nature of low-energy cutoff}

\begin{figure}\centering
\includegraphics[height=45mm]{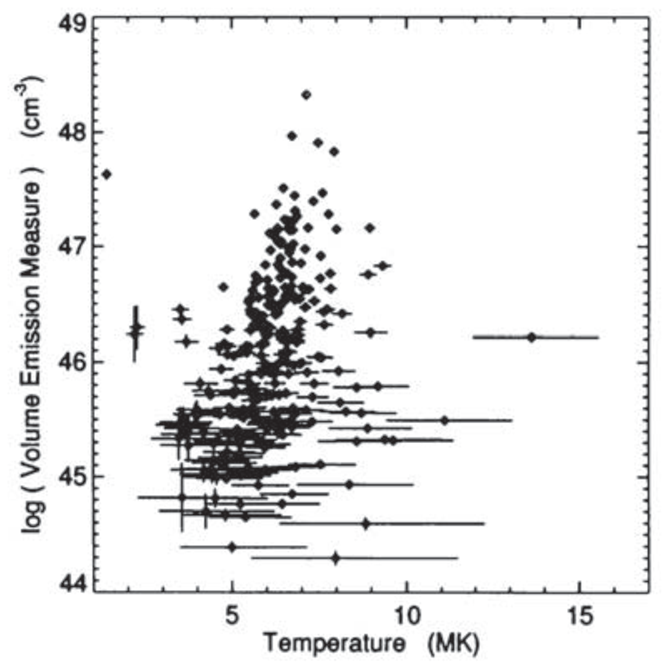}
\includegraphics[height=45mm]{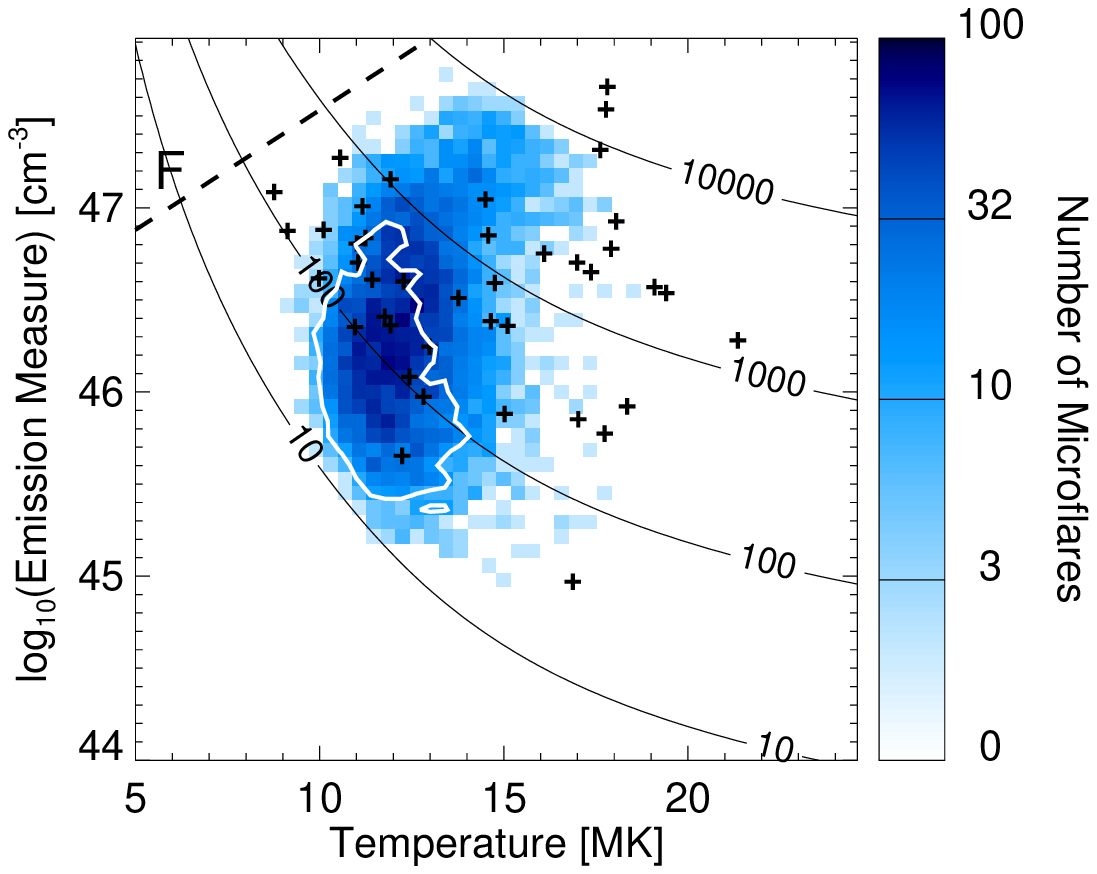}\\
\includegraphics[height=45mm]{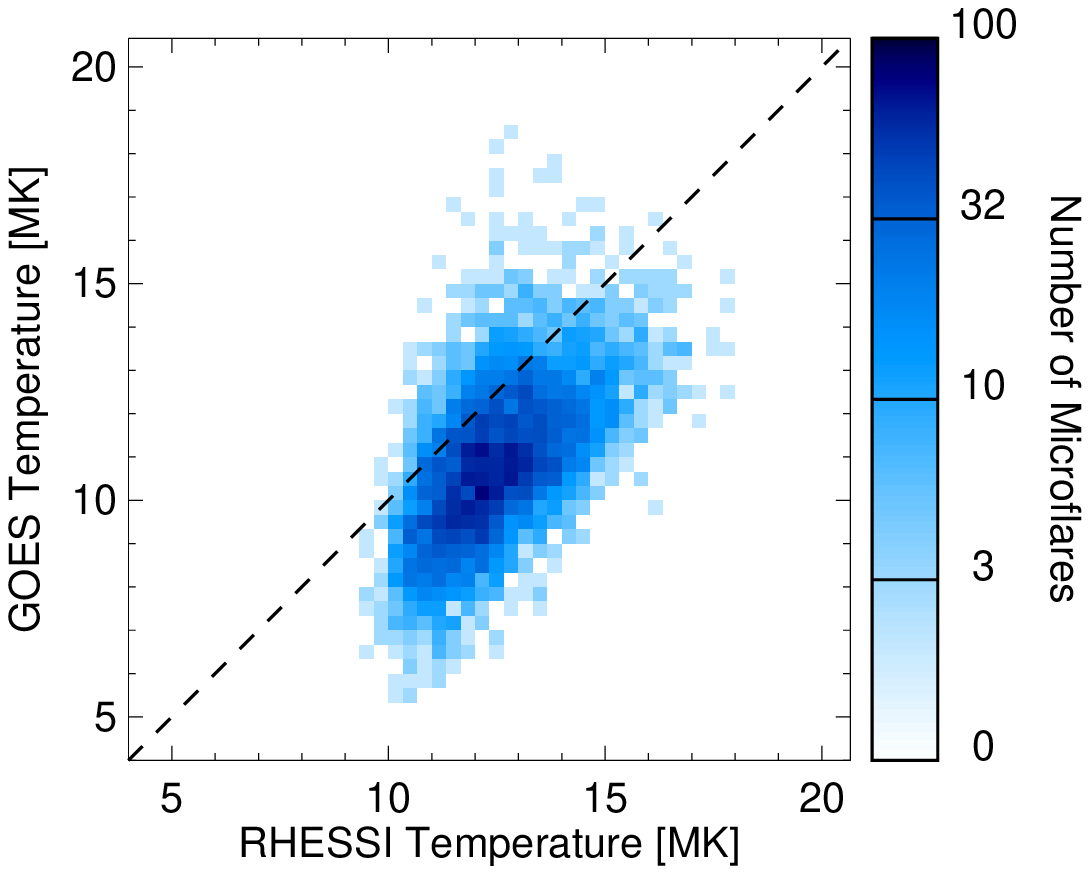}\\
\caption{\label{fig:hannah_emvst1}Thermal parameter surveys of microflares.
(\emph{Top left}) 291 \emph{Yohkoh}/SXT microflares \citep{shimizu1995} (his
``active region transient brightenings''). Reproduced by permission of the PASJ.
(\emph{Top right}) 9161 \textit{RHESSI} microflares \citep{hannah2008}. The numbered
solid lines show expected 4-8~keV \textit{RHESSI} counting-rate levels as a function of
model T and EM, the white contour shows the events with \textit{GOES} T $< 10$~MK
from the bottom panel and the dashed line (F) the relationship between
emission measure and temperature found by \cite{1996ApJ...460.1034F}; see
Figure~\ref{fig:hannah_emvst}. Note that neither of these studies individually
confirms this correlation, and that the two data sets are almost disjoint.
(\emph{Bottom}) The microflare temperature derived using both \textit{RHESSI} and
\textit{GOES} for 6740 microflares \citep{hannah2008}. Reproduced by permission of
the AAS.}
\end{figure}
\index{active-region transient brightenings!illustration}
\index{microflares!energy distributions!illustration}
\index{satellites!Yohkoh@\textit{Yohkoh}!SXT}
\index{Yohkoh@\textit{Yohkoh}!Soft X-ray Telescope (SXT)}

\subsubsection{Thermal}\label{sec:hannah_thermal}
\index{microflares!thermal spectrum}

For the basic parameters of the thermal distribution, the present standard
approach is to make an isothermal fit to a model spectrum derived from the
Chianti\index{Chianti} atomic-physics database \citep{2006ApJS..166..421L}, using standard
assumptions about abundances and ionization states.\index{ionization state}\index{abundances}
The isothermal fit determines an effective temperature~$T$ and emission measure~$n^2V$.
Figure~\ref{fig:hannah_emvst1} shows comparable regions of this parameter
space for microflares observed with the \emph{Yohkoh}/SXT \index{satellites!Yohkoh@\textit{Yohkoh}!SXT} grazing-incidence
telescope, at energies below about 2~keV \citep{shimizu1995}, top left, and by
\textit{RHESSI} at energies around $6-12$~keV \citep{hannah2008}, top right. The two
samples, though taken at different times, arguably represent the same class of
events, and yet the sets of points are almost disjoint. This illustrates the effects of
experimental bias\index{experimental bias!DEM weighting}, 
in that the isothermal approximation made for each
instrument will produce different weightings of the full DEM (differential
emission measure) distribution. This can be further seen in the bottom panel of
Figure \ref{fig:hannah_emvst1} where the temperature has been derived for the
same microflares at the same time using \textit{RHESSI} and \textit{GOES} separately. 
Clearly each instrument is responding to different parts of a DEM distribution.

\begin{figure}\centering
\includegraphics[height=70mm]{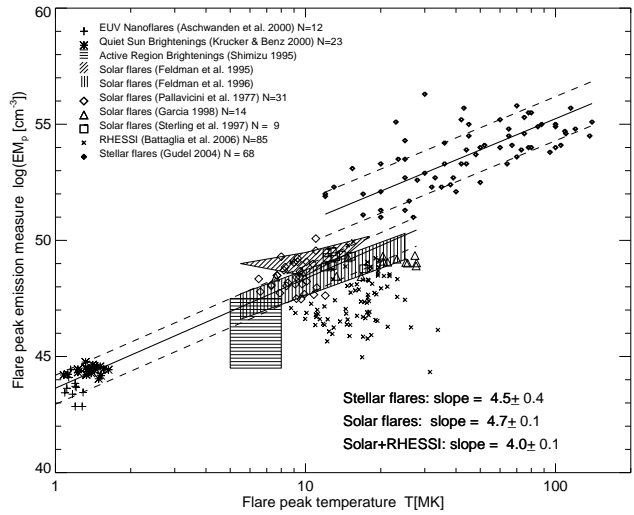}
\caption{\label{fig:hannah_emvst} Thermal fits from various surveys showing
the correlation of emission measure (log scale) with temperature of solar flare
observations compared with observations of stellar
flares and with quiet-Sun brightenings, from the sources shown
\citep{2008ApJ...672..659A}. Reproduced by permission of the AAS.}
\end{figure}
\index{emission measure!correlation with temperature!illustration}

Extending this to large flares, the thermal parameters continue to correlate in
the sense that brighter flares have higher temperatures, but with a relatively
slow growth of temperature with emission measure. \citet{1996ApJ...460.1034F}
found it to extend over three decades of flare magnitude and a factor of five in
temperature (5~to 25~MK) when using \textit{GOES} emission measure and \textit{Yohkoh}/BCS\footnote{Bent (or Bragg) Crystal Spectrometer.}
\index{satellites!Yohkoh@\textit{Yohkoh}!BCS}
peak temperature estimates of 868 A- to X-class flares. But this correlation cleanly
misses the \textit{RHESSI} microflare points, as shown by the dashed line in
Figure~\ref{fig:hannah_emvst1}; these tend to be much hotter than predicted.
The fit misses the SXT parameter space in a similar way. We presume these two
samples to represent the same physical objects, but with different observational
bias. \citet{2003AdSpR..32.1051K} found a similar correlation to
\citet{1996ApJ...460.1034F} using \textit{GOES} emission measure and temperature of
89 B- to X-class events. Analysis of super-hot flares ($>$30 MK) with \textit{RHESSI},
37 M- and X-class flares, is able to extend parameter space up to 50~MK.
\index{emission measure!correlation with temperature!RHESSI@\textit{RHESSI}}
\index{emission measure!correlation with temperature!GOES@\textit{GOES}}
The \textit{GOES} emission of these
events still correlates in a similar manner to temperature as in the
\citet{1996ApJ...460.1034F} survey, but it produces a somewhat flatter correlation in
linear-log space \citep{2010caspi}.

For individual studies over narrow ranges, there appears to be a poor correlation
between temperature and emission measure, shown in
Figure~\ref{fig:hannah_emvst}. 
A possible correlation appears if many studies
over a wider range are considered \citep{2008ApJ...672..659A} incorporating even
stellar flares and quiet-Sun brightenings. 
\index{emission measure!correlation with temperature!broad parameter ranges}
Taken together one seemingly obtains
a definite correlation from sample to sample, if not within a given data set.
However, this may be force-fitting a single concept to different things; a single
power-law fit to all of the points would be describable approximately by $EM
\propto T^7$, but if one ignores the quiet-Sun events one might prefer  a much
steeper relationship such as $EM \propto T^{15}$. Given the strong systematic
biases among the different kinds of observation represented, perhaps linked
only by the word ``flare,'' it is no doubt premature to draw any strong
conclusions and caution is required when making such comparisons.

\begin{figure}\centering
\includegraphics[height=60mm]{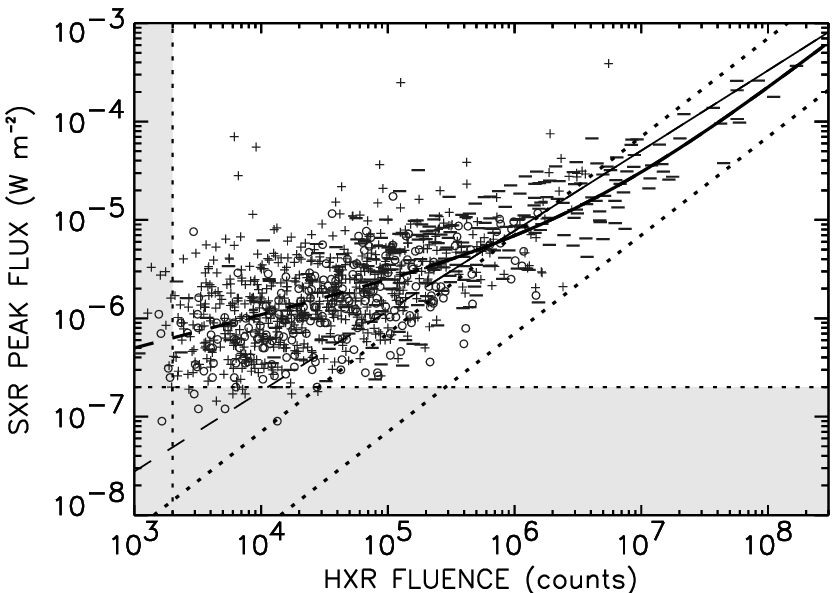}\\
\includegraphics[height=45mm]{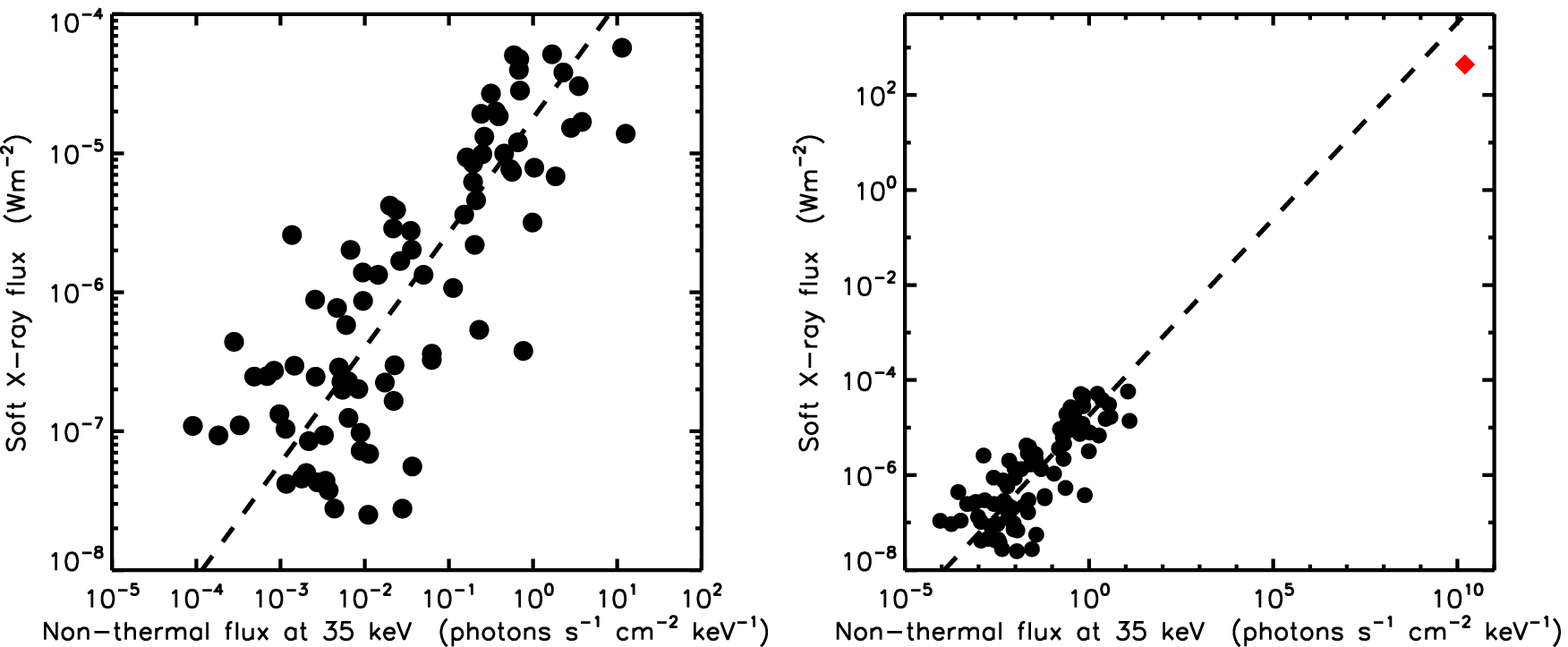}
\caption{
\label{fig:hannah_thvsnth}Comparisons of thermal and nonthermal fluxes for
ordinary flares. \emph{Upper:} hackground-subtracted SXR peak flux from \textit{GOES}
vs. HXR fluence from HXRBS \citep{2002A&A...392..699V}. The solid lines show
various functional fits, the dashed lines show proportionality, with extension
to smaller fluence, and the grey areas indicate regions below instrumental threshold.
The different plot symbols indicate the time difference between SXR and HXR
peaks. \emph{Lower:} a representative sample of \textit{RHESSI} flares
\citep{2005A&A...439..737B}, plus the same sample but including the enormous
stellar flare observed by \cite{2007ApJ...654.1052O} (the red point). Again, the
dashed line shows proportionality.}
\end{figure}

\subsubsection{Thermal-nonthermal relationships}
\index{microflares!thermal/nonthermal relationship}

For major flares we note that the impulsive-phase nonthermal signatures often have
a simple physical relationship with the gradual thermal signatures, namely the
Neupert effect where the time integral of the HXR (and microwave) emission is
empirically seen to match the SXR time profile
\citep{1968ApJ...153L..59N,1991BAAS...23R1064H,1993SoPh..146..177D}. 
\index{Neupert effect!in microflares}
Does this relationship extend into the microflare domain as well?
Figure~\ref{fig:hannah_thvsnth} shows how well the thermal and nonthermal
signatures relate. 
\index{gradual phase!in microflares}
The \textit{GOES}\index{satellites!GOES@\textit{GOES}} peak flux and HXRBS (Hard X-Ray Burst Spectrometer
on the \textit{Solar Maximum Mission}, \textit{SMM})\index{satellites!SMM@\textit{SMM}!HXRBS} fluences show a large scatter for weaker
events, but for the better-observed large energetic events approach a
proportional relationship with little scatter \citep{2002A&A...392..699V}. One
would expect greater scatter in the fainter events purely due to selection effects,
instrumental sensitivity and analysis procedure (e.g., background subtraction
performed).

\begin{figure}
\centering\sidecaption
\includegraphics[height=55mm]{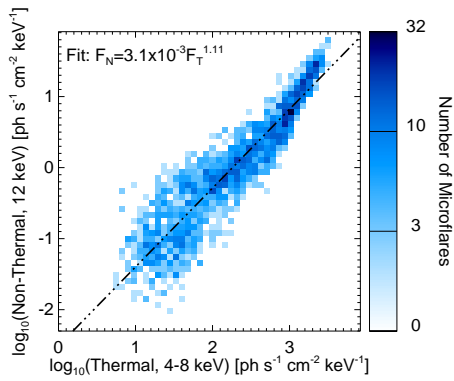}
\caption{
\label{fig:hannah_thvsnth2} Thermal/nothermal correlation for 4236 \textit{RHESSI}
microflares \citep{hannah2008}. The nonthermal flux is the estimated 12~keV
emission using the fitted nonthermal model.\index{visibilities} 
The thermal flux is found from fitting the X-ray visibilities over $4-8$~keV. 
Note that the axes are inverted relative
to those of Figure~\ref{fig:hannah_thvsnth}, but that the line of proportionality
is the same. Reproduced by permission of the AAS.}
\end{figure}

For \textit{RHESSI}, a similar effect can be observed
\citep{2005A&A...439..737B,2007SoPh..246..339S}, as shown in the lower panel of
Figure~\ref{fig:hannah_thvsnth}), although here the fluxes are directly compared, rather than the
SXR flux and HXR fluence. 
Also shown is one of the most energetic stellar flares
observed \citep{2007ApJ...654.1052O}, over six orders of magnitude brighter in SXRs
than the largest solar flare. 
Here again the SXR and HXR fluxes scaled together as in the ``big flare syndrome.''
\index{big-flare syndrome!microflares}
\index{syndromes!big-flare!microflares}
This relationship appears to extend into the microflare domain.
Figure~\ref{fig:hannah_thvsnth2} shows that it does, using the large sample of
\textit{RHESSI} microflares studied by \cite{hannah2008} and \cite{christe2008}. 
Note that the deviation from the trend line for the largest events is instrumental and due to
increased detector deadtime before \textit{RHESSI}'s thin shutter deploys. 
\index{RHESSI@\textit{RHESSI}!attenuating shutters!and deadtime}
In general, Figure~\ref{fig:hannah_thvsnth2} provides additional strong evidence that the
microflares simply represent an extension, to lower energies, of the same
physical processes at work in flares. \cite{asch2007} has pointed out that such a
relationship depends upon the scaling of the event environment; at some point
we would expect the weakest and the most powerful events to show some kind of
deviation from this behavior -- for example, low-altitude loops might have
shorter coronal cooling times and therefore depart 
systematically\index{flare frequency distributions!and natural limits} from the
observed thermal/nonthermal correlation.

Detailed quantitative analysis of flare thermal and nonthermal emission suggests
that the underlying mechanisms are more complicated than empirically shown via
the Neupert effect.
\index{Neupert effect}
\index{emission measure!and Neupert effect}
\index{satellites!Yohkoh@\textit{Yohkoh}!SXT}
\index{satellites!Yohkoh@\textit{Yohkoh}!SXT} 
\index{satellites!Yohkoh@\textit{Yohkoh}} 
The differential emission measures (DEMs) for 80 flares were
studied with the \textit{Yohkoh}/SXT and BCS instruments
with the finding that the high-temperature plasma
($>$16.5~MK) is more likely to demonstrate the Neupert effect\index{Neupert effect} than lower
temperature plasma \citep{1999ApJ...514..472M}.
Given that the Neupert effect is
thought to show that the accelerated electrons are responsible for heating the
chromospheric plasma, \citet{2005ApJ...621..482V} investigated the similarity of
the power in the electron beam\index{electrons!beams} compared to the power required to produce the observed SXRs.
They expected the powers to be better correlated than the HXR
fluence and SXR flux but found a similar correlation. 
One possibility for this could
be that the heating of SXR-emitting plasma is not solely due to the electron beam
that produces the HXR emission.

\subsection{Energy distributions}\label{sec:hannah_engdist}
\index{flare frequency distributions}

As previously introduced, the thermal and nonthermal energies of flares cannot be
determined directly from observations but have to be inferred using model
assumptions.  The ambiguities of these models and the errors and bias on the
observational parameters result in large uncertainties in the energy, and hence
in the energy distribution.

The thermal energy $U_\mr{th}$ can be estimated via
\begin{equation}\label{eq:therm}
U_\mr{th}=3n_\mr{e}k_\mr{B}T V=3k_\mr{B}T\sqrt{EM\cdot V_\mathrm{obs}f},
\end{equation}
\noindent where $V$ is the volume of the emitting plasma and $f$ is the filling factor
(the ratio of the actual volume to the measured value, such that $V=f
V_\mathrm{obs}$, where $n_\mr{e}$ is the plasma density, $T$ is the temperature, and
$EM=n_\mr{e}^2V$ is the emission measure). 
\index{filling factor!distribution functions}
The temperature and emission
measure can be found through fitting an isothermal model spectrum to either the
flare spectrum or the ratio of images in multiple wavelengths. The volume is
estimated from the 2-D area of emission from the images, with some model
assumption to convert to a volume. The thermal energy (given by Equation~\ref{eq:therm})
is an upper limit estimate over the time of the observed spectrum
and images, since a filling factor of $f=1$ is normally assumed \cite[see, e.g.,][]{chapter2}, though in reality it
could be considerably smaller, $10^{-3}$ to $10^{-2}$
\citep{1997ApJ...478..799C,2000AdSpR..25.1833T}.

The power in the nonthermal electrons of energies above $E_\mr{C}$ (in keV) is
estimated via
\begin{equation}\label{eq:pow} P_\mr{N}(>E_\mr{C})=9.5\times 10^{24}
\gamma^2 (\gamma-1) \beta \left(\gamma -0.5,1.5 \right)
E_\mr{C}^{1-\gamma}I_\mr{0}\quad \mr{erg~s}^{-1},
\end{equation}
\noindent where $\gamma$ and $I_\mr{0}$ are the index and normalization of the
power law in the observed photon spectrum (in units of photon spectral flux) and
$\beta(m,n)$ is the beta function \citep{1971SoPh...18..489B,1974SSRv...16..189L}.
To estimate the energy in the nonthermal electrons
\index{electrons!total energy} $U_\mr{n}$ 
the total duration of the HXR emission needs to be known.
\index{flare models!thick-target}
\index{electrons!beams}
The standard ``cold thick-target model'' model assumes a chromospheric thick target in
which a beam of electrons stops, such that the thermal energy of the ambient electrons is negligible compared to the energies of the electrons emitting the HXR.

\textit{RHESSI} crucially allows the thermal and nonthermal energies to be determined in
a flare using a single instrument, due to the energy range it covers and to its imaging
and spectroscopic capabilities. Studies of several large flares (C-, M- and X-class)
have found that the thermal and nonthermal energies are the same, to within an
order of magnitude \citep[e.g.,][]{2003ApJ...595L..97H,2005A&A...435..743S}. 
This suggests that the conversion of the energy in the accelerated particles into thermal
plasma energy is highly efficient. 
For C- and M-class events this is typically about $10^{30}$
erg \citep{2005A&A...435..743S}, and at minimum about $10^{31}$ erg for large
X-class events \citep{2003ApJ...595L..97H}.  
Another study found that the power
required for the white-light emission\index{flare types!white light!energy}\index{white-light flares!and HXR}
was comparable to the electron beam power required to produce the HXR emission, but only if the cutoff was less than 25 keV
\citep{2007ApJ...656.1187F}.

For microflares, the situation is more complicated, given the difficulty in
determining the properties of the photon spectrum and the uncertainty this leads
to in estimating the break energy.
\index{microflares!energy estimates}
The first detailed analysis on individual \textit{RHESSI}
microflares (A- and B-class flares) found that the steepness of their spectra results in
considerable power in the these low-energy electrons, but only estimated the
energy above 25 keV at $10^{26}-10^{27}$~erg \citep{2002SoPh..210..445K}. 
In a larger study of microflares, where the observed break found in the photon spectrum was
used as the cutoff energy, the nonthermal energy ranged over $10^{28-31}$~erg
\citep{2007SoPh..246..339S}. 
\index{low-energy cutoff!in microflare hard X-ray spectra}
This gives an overestimate in the energy since
$\epsilon_B$$< E_C$, and using slightly larger values for $E_C$ produced typical
nonthermal energies of about $10^{28}$ erg, similar to the thermal energies found.
In the largest \textit{RHESSI} microflare study \citep{hannah2008} the thermal and
nonthermal energies at about the time of peak emission over $6-12$~keV were found to
be similar, to within an order of magnitude, once an empirical correction factor had
been used to convert the measured $\epsilon_B$ to $E_C$. 
This factor was found
by fitting a broken power laws to model thick-target spectra with a range of indices
and low energy cutoffs. Further discussion on the calculation of flare energy and
the associated issues is given in \citet{chapter3}.

The distribution of these \textit{RHESSI} microflare values is shown in Figure~\ref{fig:hannah_eng},
 in comparison to previous HXR nonthermal energy
distributions \citep{crosby1993,lin2001}, microflares from the SXR thermal
distribution \citep{shimizu1995} and the thermal distribution of EUV nanoflares.
\citep{parnell2000,asch2000,benz_krucker2002}
\index{flare frequency distributions!microflare comparison}
\index{flare frequency distributions!nonthermal} 
Although such a figure nicely shows the scaling of nanoflares\index{nanoflares!selection effects} through microflares to large flares, and is crucial for determining coronal heating, it is dangerously deceptive. 
Each of these distributions was found for a different type of event, using various instruments
and for different periods during the solar cycle, and so each will be affected by
different types of selection effect and biases. For instance, the SXT energies
\citep{shimizu1995} are from 291 brightenings in one active region over five days in
1992 August, whereas the \textit{RHESSI} microflare energies \citep{hannah2008} are from
9161 events taken over five years, March 2002 to March 2007. 
This is a comparison of probably the same type of event but even then they are still observed at different wavelengths and over different time periods. Extending this comparison to EUV
nanoflares\index{nanoflares!EUV}\index{flare frequency distributions!nanoflare range} is very difficult and may not even be appropriate given that we do not know if these are similar events or completely distinct physical processes.
When looking at the actual values of $\alpha$ that these distributions provide, we
see the striking observational result that flare peak flux distributions are power
laws flatter than $\alpha = 2$, a result that goes back at least to
\cite{1956PASJ....8..173A}.
This appears to exclude nanoflares\index{flare frequency distributions!nanoflares!inconsistency} as the source of coronal heating \citep{1991SoPh..133..357H}. 
On the other hand, distributions that
attempt to reflect the event energies themselves have often found steeper
spectra that are consistent with Parker's\index{Parker, E. N.} idea. 
\citet{parnell2000} found that $\alpha$ ranged from 2.1 to 2.6 for different model assumptions but was
consistently in the $>$2 range. \citet{1998ApJ...501L.213K} found that $\alpha$ was
between 2.3 and 2.6 but later showed that these values and those from other EUV
nanoflare studies were highly dependent on the flare selection criteria
\citep{benz_krucker2002}. 

So, could selection effects be responsible for this
discrepancy between nanoflare and larger flare distributions? 
Attempts have been made to recover the intrinsic distributions from the observationally derived
ones \citep{asch2000,asch_parnell2002} using a technique similar to that applied
to Malmquist bias\index{Malmquist bias}, a threshold selection effect that biases galaxy number counts in cosmology \citep{1990A&A...237..275H,1994ApJS...92....1W}.
In this situation, the counting statistics are biased because the brighter, more distant galaxies are more likely to be counted than the fainter ones.  
This Malmquist-bias procedure has been used on the thermal energies
derived for EUV nanoflares\index{nanoflares!EUV}; it takes the means and covariances of the observed
data parameters and iterates them back to the intrinsic unbiased values. To do this
requires model assumptions about how the parameters required to calculate the
thermal energy, in Equation \ref{eq:therm}, relate to each other, so that the
probability distribution of the intrinsic distribution can be analytically described.
However, determining how these parameters scale is, in itself, subject to biases and
selection effects, as discussed in Section \ref{sec:hannah_thermal}. The analysis
procedure may then only adjust the scalings and not test whether the parameters
actually relate in such a manner. 
Attempting to recover the intrinsic distributions,
free of instrumental and selection effect biases, should certainly be the priority of
any study deriving flare distributions, but in practice this does not readily appear to be
practicable given that the parameter of greatest interest, the energy, is not directly
obtainable from the existing observations.
An alternative approach of determining flare energy, which is much less prone to these
effects, is to measure the luminous component directly via the resulting change in the total solar irradiance
(TSI).\index{TSI}
This has been done for the X17 event SOL2003-10-28T11:10 
\index{flare (individual)!SOL2003-10-28T11:10 (X17.2)!TSI signature} 
using the TIM (Total Irradiance Monitor) instrument on the \textit{Solar Irradiance and Climate Experiment (SORCE)}
\index{satellites!SORCE@\textit{SORCE}!TIM} 
\citep{2005SoPh..230..129K}, which is sensitive to the solar
emission from X-rays to far infrared. 
The estimated total luminous energy for this flare was
approximately $5\times10^{32}$~erg \citep{2004GeoRL..3110802W}. 
This technique is only appropriate at present
for the largest flares, because of competition from other sources of solar variation.

Given the evidence so far, it does not appear that flare-like events can heat the
quiescent corona. 
Moreover it seems that making the comparison of active-region
and quiet-corona flare events is unwise given that these transient events
have different populations in appear in different physical environments
\citep{benz_krucker2002}.

\section{Conclusions \& discussion}\label{sec:hannah_cons}

The \textit{RHESSI} data have allowed us to study microflares effectively while using the
same instrumentation for major events. The statistical study of solar flares has
great importance in understanding the underlying processes involved in the
energy release and subsequent emission we observe in the solar corona. Often,
emphasis is placed on detailed multi-wavelength studies of individual events.
Such studies have their merits in revealing insights to the processes in these
events. However without being able to place the flare in context of other events, it
is very difficult to determine whether it is typical or unusual behavior that is
being studied. Only by studying events in large numbers with a minimum of
selection bias can one really approach an understanding of the general physics.

The discovery that ``ordinary'' flare physics extends down to the tiniest events
observable by \textit{RHESSI} (or by \textit{GOES}) allows us to conclude that such events do not
explain coronal heating. Even the smallest events \textit{RHESSI} observes are in the
active regions and are flare-like, distinguishable as individual events. The
nanoflare hypothesis instead requires an apparently continuous flare population
that is many orders of magnitude smaller. They also continue to show the flat
peak-flux scaling that puts most of the energy in the most powerful events, rather
than the weakest ones. In a sense, this conclusion simply confirms the appearance
of the \textit{GOES} data -- sometimes flare-dominated, and sometimes showing steady
emission. This is inconsistent with the universality of the flat flare power-law distribution. 
Of course, numerous individually unobservable nanoflares could create the
apparently steady emissions.

To make further progress in this field, three crucial things have to happen. Firstly,
we need spacecraft that have a higher sensitivity, lower background and wider
dynamic range than \textit{RHESSI} while maintaining the energy range covered across
both imaging and spectroscopy. This would allow the faintest events in active
regions and the quiet Sun, as well as the fainter emission components of large
flares, to be confidently analyzed. One such suitable implementation would be a
HXR focusing-optics telescope dedicated to solar observations. 
Second, we must understand better how biased our observations are, and how we
can obtain the intrinsic unbiased physics from our observations. 
This will require a third advance, more sophisticated modeling of these faint events to
match improvements in the data and their implications.

\begin{acknowledgements}
The authors thank the editor (B. R. Dennis) and the two referees for
providing many constructive comments that have greatly improved this review.
IGH would also like to thank the Glasgow SSH (Solar Self Help) group for their
comments. This work is financially supported through NASA contract NAS
5-98033, an STFC rolling grant, and by the European Commission through the SOLAIRE
Network (MTRN-CT-2006-035484), and all are gratefully acknowledged. JK
acknowledges support from Grant 205/06/P135 of the GA CR and the research plan
AV0Z10030501.
\end{acknowledgements}

\appendix
\section{Determining distribution parameters}\label{app:hannah_fitpowlaw}

For a power-law distribution (Equation~\ref{eqn:hannah_powlawdef}), the
standard way to determine the index $\alpha$ is to perform a linear fit to the
log-log histogram of the data. 
\index{distributions!power law}
However this is a highly subjective approach as
there is considerable choice as to the ``best'' bin width and fitting method. An
alternative and more objective approach is to estimate the power-law index using
the maximum-likelihood method \citep{1970ApJ...162..405C,bai1993}. This
approach leads to a remarkably simple calculation on the sample to determine the
index above some chosen threshold:

\begin{equation} \alpha_\mr{m}=\frac{N}{\sum_{i=0}^N \ln{(U_i/U_0)}}+1.
\end{equation}

\noindent In this example, the energies $U$ are used, where $N$ is the total
number of events and $U_i$ is the energy of the $i^{th}$ event normalized by a
threshold energy of $U_0$. The error in this most likely value of $\alpha_\mr{m}$
can be estimated \citep{wheatland2004} as

\begin{equation}
\sigma_\alpha=(\alpha_\mr{m}-1)N^{-1/2}.
\end{equation}

\noindent The observed distribution, however, is often affected by instrumental
and selection effects, resulting in a deviation from a power law, so fitting a
power law alone would be unwise. A common problem is that the smallest events
are hard to detect and analyze successfully, resulting in a flattening of the
power-law for these missing smallest events. To fit this observed biased
distribution a skew-Laplace distribution can be used instead of a single power law
\citep{parnell2000}. This fits the desired power-law distribution for the larger events,
but below some critical value, a power law with a different index is fitted to the under-reported
smallest events. 
The determination of the parameters of this skew-Laplace
distribution again can be found using a maximum likelihood estimation method
\citep{parnell2000}.
\index{distributions!skew-Laplace}

Another possible distribution to describe the observed sample of data is the
Weibull distribution \citep{parnell2002}. 
\index{distributions!Weibull}
This distribution has the form

\begin{equation}
f(x;\kappa,\xi)=C \left (x/\xi \right )^{(\kappa-1)}\exp[{-(x/\xi)^\kappa}],
\end{equation}

\noindent where $\kappa$ is the shape parameter and $\xi$ is the scale
parameter. For a shape parameter $\kappa<$1, the resulting distribution is similar
to a power law, but it turns over at the smallest and largest events.
This can then represent deficiencies in the smallest and largest events. Again, the
under-reporting of the smallest events is likely to arise from instrumental
sensitivity and selection effect bias. 
The largest events may be missing due to the
limited dynamic range of the instrument, if these events saturate the detector; or,
there might actually be fewer of these events if a critical physical upper limit is
being reached (for instance maximum energy available in an active region
\citep{1997ApJ...475..338K}. 
Such a feature is consistent with the predictions of
avalanche models \citep{1993ApJ...412..841L}, as discussed in Section
\ref{sec:hannah_whyapowlaw}. 
\index{flare models!avalanche}
The parameters of this Weibull distribution can
be determined again using the maximum likelihood method \citep{parnell2002}.

As there are several distributions that could successfully fit the data, a statistical
test is needed to determine which is best. Such a test is the Kolmogorov-Smirnov
statistic (KS), which is the maximum difference between the cumulative
distribution function (CDF) and the empirical distribution function (EDF)
\citep{1992nrca.book.....P}. For a data set of, energies $U$, the CDF for the $i^{th}$
energy $U_i$ is the integral of the PDF, using the fitted parameters, to
$U_i$. 
The EDF is derived from the observed/calculated parameters and in this
example is the number of events with energy less than or equal to $U_i$, which
turns out to be $(i-1/2)/N$. The KS statistic then provides a measure of the
significance level of each distribution \citep{1992nrca.book.....P}. 
Plotting CDF versus EDF provides a graphical way of determining how consistently the data 
belong to the chosen distribution. 
A graph similar to the familiar histogram can
be obtained by plotting $1 -$~CDF and EDF against the event parameters
(i.e., energy).

A detailed example of using the maximum likelihood method to determine the
parameters, and testing the goodness of the fit using the KS statistic for
power-law and Weibull distributions in the solar context, is given in
\citet{parnell2002}.


\printindex

\end{document}